%% file: paper_arxiv.tex
\documentclass[12pt]{article}
\include{_preamble}
\onecolumn

\makeatletter
\@ifclassloaded{oup-authoring-template}{%
    \journaltitle{Journal of the Royal Statistical Society, Series B}
    \DOI{DOI HERE}
    \copyrightyear{XXXX}
    \pubyear{XXXX}
    \access{Advance Access Publication Date: Day Month Year}
    \appnotes{Original article}
    \firstpage{1}

    \title[Modified treatment policies under network interference]{\titlepaper}
    \author[1]{Salvador V.~Balkus}
    \author[2]{Scott W.~Delaney}
    \author[1,$\ast$]{Nima S.~Hejazi}
    \authormark{Balkus, Delaney, and Hejazi}

    \address[1]{\orgdiv{Department of Biostatistics}}
    \address[2]{\orgdiv{Department of Environmental Health},
    \orgname{Harvard T.H.~Chan School of Public Health},
    \orgaddress{\street{677 Huntington Avenue},
    \postcode{Boston}, \state{MA 02115}, \country{USA}}}
    \corresp[$\ast$]{Correspondence: Nima S.~Hejazi, Department of
    Biostatistics, Harvard T.H.~Chan School of Public Health,\\677 Huntington
    Avenue, 1-411, Boston, MA 02115, USA.
    Email: \texttt{nhejazi@hsph.harvard.edu}}

    \received{Date}{0}{Year}
    \revised{Date}{0}{Year}
    \accepted{Date}{0}{Year}

    \abstract{\abstracttext}
    \keywords{causal inference, stable unit treatment value assumption, network
      dependence, semi-parametric estimation, efficient influence function,
      causal machine learning}
}{%
  \author{\authorlist}
  \title{\titlepaper}
}
\makeatother

\zexternaldocument*[supp:]{sm}
\begin{document}
\maketitle
\makeabstract % uncomment for arxiv
% \footnotetext{\firstauth}

\section{Introduction}\label{section:introduction}

Scientists frequently seek to understand the causal effect of a given
policy on an outcome $Y$ in a target population. Such policies commonly involve
\jrssb{modifying the natural value of some continuous
treatment or exposure $A$; that is, the value $A$ would have taken had its distribution been left unaltered \citep{Sarvet2025}}. Often, the
scientist's ideal goal is to answer the question, ``how
would the population mean of $Y$ change if the natural value of $A$ were
increased or decreased?''

\textbf{Modified treatment policies} (MTPs)~\citep{robins2004,
  munozPopulationInterventionCausal2012,
haneuseEstimationEffectInterventions2013, young2014identification} are a class
of interventions that define causal estimands well-suited for formulating such
counterfactual questions about continuous exposures. An MTP can answer the
question, ``how much would the average value of $Y$ have changed had the
natural value of $A$ been increased or decreased by an increment $\delta$?''
where $\delta$ is chosen by the investigator. Hence, the MTP provides a
\textit{causally interpretable generalization} of the commonly applied
procedure of estimating a regression coefficient. In fact, under specific
structural assumptions, the effect of applying the MTP $A + 1$ to every study
unit is equivalent to the coefficient for the exposure $A$ in a main terms
linear regression. Hence, though rarely mentioned explicitly by name, the
causal effect of an MTP is a popular target estimand in studies that aim to
evaluate policy effects on a well-defined population.

The causal effect of an MTP carries several key advantages over
alternative estimands. Unlike the causal dose-response curve, the MTP effect
is identified even in settings where it may not be feasible---or, indeed,
sensible---to deterministically set each unit's exposure to the same level.
Consequently, the MTP effect is identified even when certain exposure levels
may suffer from non-overlap among sub-populations. Furthermore, the MTP
effect estimand admits standard semi-parametric efficient estimators,
facilitating the use of machine learning to flexibly capture nonlinear
relationships and thus allowing the investigator to mitigate the risks posed by
model misspecification.

Like other common causal inference methods, a key assumption of the
classical MTP framework is \textit{non-interference}---that one unit's
exposure does not impact any other units' outcome~\citep{cox1958planning,
rubin1980}. This critical assumption cannot be made
when, for example, study units correspond to geographic areas, such as counties
or ZIP code areas. Here, the population units may move around; thus,
assigning an exposure policy to a given county would affect not only those
residing in that geographic region but also those commuting to and from said
region. In general, ignoring interference between units invalidates
identification and overtly introduces bias into analytic results
\citep{halloranDependentHappeningsRecent2016}. Despite these challenges, data
from studies that involve interference between study units can be exceedingly
useful, even critical, to advancing science and policy in many settings of
current interest. For example, environmental epidemiologists routinely
leverage observational spatial data to study the effects of pollution on human
health~\citep{elliottSpatialEpidemiologyCurrent2004,
reichReviewSpatialCausal2021a, morrisonDefiningSpatialEpidemiology}, a setting
in which randomization is impractical and unethical, and where the
``shifts in distribution'' measured by MTPs are of interest~\citep{Tec2024}.

Owing to its straightforward interpretability and the non-restrictive
conditions for its identification, especially when compared to other causal
estimands for continuous exposures, it is useful to deepen understanding of how
to apply the MTP framework even in settings when non-interference cannot be
assumed. The present work develops a framework for both identifying and
efficiently estimating an MTP's causal effect under network interference.
This allows investigators to obtain inference about the causal effect of a
continuous exposure while both accounting for network interference between
units and incorporating flexible regression procedures in the estimation
process.

\textbf{Contributions.} We present several contributions to the
literature on causal inference under network interference. Firstly, we
introduce the concept of an \textit{induced MTP}, a new type of intervention
that identifies the causal effect of an MTP when network interference is
present. This intervention accounts for how the application of an MTP to a
given unit affects its neighbors in the known (or assumed) network structure.
Secondly, applying the ``coarea formula'' from measure-theoretic calculus, we
provide a novel identification result for the causal effect of an induced MTP
in the network setting. Finally, we develop semi-parametric efficient
estimators for the causal effect of an induced MTP by applying and building on
the prior theoretical contributions of~\cite{ogburnCausalInferenceSocial2022}.
Specifically, we resolve several  of their methodological challenges, including (1) using cross-fitted
machine learning for nuisance parameter estimation in place of restrictive
parametric regression strategies; (2) deriving more tractable forms of the
relevant nuisance parameters, which can be reliably estimated using standard
regression tools; (3) eliminating reliance on computationally-intensive Monte
Carlo procedures for estimation and inference; and (4) obtaining consistent
variance estimators.

\textbf{Outline.} Section~\ref{section:background} reviews MTPs and causal
inference under interference. Section~\ref{section:methods} describes the
induced MTP, including identification and estimation of causal effects. Section~\ref{section:sim-results} reports numerical
results verifying the proposed methodology, while
Section~\ref{section:data-analysis} discusses an illustrative data analysis in
our motivating applied science context. We conclude and discuss future
directions in Section~\ref{section:discussion}.

\section{Background}\label{section:background}

Throughout, we let capital bold letters denote random $n$-vectors; for
instance, $\Yb = (Y_1, \ldots, Y_n)$. Consider data $\Ob = (\Lb, \Ab, \Yb) \sim
\Pf \in \M$, where $\Pf$ is the true and unknown data-generating distribution
of $\Ob$ and $\M$ is a non-parametric statistical model (that is, a set of
candidate data-generating probability distributions) that places no
restrictions on the data-generating distribution. In principle, $\M$ may be
restricted to incorporate any available real-world knowledge about the system
under study. Let $O_i = (L_i, A_i, Y_i)$ represent measurements on the
i\textsuperscript{th} individual data unit, where $Y_i$ is an outcome of
interest, $A_i$ is a continuous exposure with support $\A$, and $L_i$ is a
collection of baseline (that is, pre-exposure) covariates. To ease notational
burden, we omit subscripts $i$ when referring to an arbitrary unit $i$ (that
is, $Y \equiv Y_i$ when the specific unit index is taken to be uninformative).

We will assume that the data-generating process can be expressed via a
structural causal model (SCM, \cite{pearl2000models}) encoding the temporal
ordering between variables: $\Lb$ is generated first, then $\Ab$, and finally
$\Yb$. We denote by $Y(a)$ the counterfactual random variable (or potential
outcome) generated by hypothetically intervening upon $A$ to set it to $a \in
\A$ and allowing the impact of such an intervention to propagate downstream, to
the component of the SCM that generates $Y$. Our goal is to reason about the
causal relationship between $\Ab$ and $\Yb$ in spite of the presence of
confounders $\Lb$ and network interference between units $i = 1, \ldots, n$.

\subsection{Continuous Exposures with Modified Treatment Policies}

A \textbf{Modified Treatment Policy} (MTP) is a user-specified function $d(a,
l; \delta)$ that maps the observed value $a$ of an exposure $A$ to a new value
and may itself depend on the natural (or pre-intervention) value of the
exposure~\citep{haneuseEstimationEffectInterventions2013}.
\begin{example}[Additive Shift]
For a fixed $\delta$, an \textit{additive shift} MTP may be defined as
\begin{equation}
    d(a, l; \delta) = a + \delta \ .
    \label{add-mtp}
\end{equation}
This corresponds to the scientific question, ``how much of a change in $Y$
would be caused by adding $\delta$ to the observed natural value of $a$, for all
units regardless of their stratum $l$?''
\label{additive-shift}
\end{example}
\begin{example}[Multiplicative Shift]
For a fixed $\delta$, a \textit{multiplicative shift} MTP is defined as
\begin{equation}
    d(a, l; \delta) = \delta \cdot a \ .
    \label{mult-mtp}
\end{equation}
This asks the scientific question, ``how much of a change in $Y$ would be
caused by scaling the observed natural value of $a$ by $\delta$, for all units
regardless of their stratum $l$?''
\label{multiplicative-shift}
\end{example}
\noindent
Note that in the above, $\delta$ is a fixed, user-specified parameter
specifying the magnitude of the hypothetical intervention. However, the MTP
framework is even more flexible: interventions that change depending on the values
of measured covariates $L$ are allowed.
\begin{example}[Piecewise Additive Shift]
Consider a piecewise additive function
\begin{equation}
    d(a, l; \delta) = \begin{cases}
      a + \delta\cdot l & a \in \A(l) \\
      a & \text{otherwise} \ ,
   \end{cases}
   \label{padd-mtp}
\end{equation}
which applies an intervention whose scale depends on the value of a covariate
$l$ and only occurs if the natural exposure value $a$ is within some specific
subset $\A(l) \subset \A$ of the support of $A$.
\end{example}

MTPs can be used to define scientifically relevant causal estimands for
continuous exposures. The \textit{population intervention causal effect of an
MTP}~\citep{munozPopulationInterventionCausal2012} is defined as
$\E_{\Pf}[Y(d(A, L; \delta)) - Y]$; that is, the average difference
between the outcome $Y$ that did occur under the observed natural value of
treatment $A$ and the counterfactual outcome $Y(d(A, L; \delta))$ that would
have occurred under the investigator-supplied MTP $d(A, L; \delta)$. This
estimand answers the scientific question, ``what would happen if we applied, to
the study population, a policy that modified the existing exposure according to
a rule encoded by $d(\cdot; \delta)$?''

When $\delta = 1$, the additive MTP (Example~\ref{additive-shift}) carries the
familiar interpretation often attributed to a linear regression coefficient; when $\delta = 1.01$, the multiplicative MTP (Example~\ref{multiplicative-shift})
holds the same interpretation, but for a log-transformed exposure. Hence, MTPs may be seen
as a non-parametric and causally interpretable extension of widely used
associational estimands, formalizing the problem of quantifying how the mean
counterfactual outcome would change under a shift in exposure value. MTPs allow
the investigator to specify a wide range of interpretable interventions on
continuous exposures that may be carried out in practice. The MTP framework has
gained traction for its applicability in settings involving longitudinal,
time-varying interventions \citep{diazNonparametricCausalEffects2023,
hoffman2024studying}; causal mediation analysis~\citep{diaz2020causal,
hejazi2023nonparametric}, including with time-varying mediator--confounder
feedback~\citep{gilbert2024identification}; and causal survival analysis under
competing risks~\citep{diaz2024causal}. However, no work has, to our knowledge,
extended the MTP framework to settings with dependent data characterized by
interference between units.

% MTPs provide an interpretable causal estimand for continuous treatments.
\subsection{Causal Inference Under Interference}

Interference occurs when, for a given unit $i$, the outcome of interest $Y_i$
depends not only on its own assigned exposure $A_i$ but also upon the exposure
$A_j$ of at least one other unit ($j \neq i$). Formally, we say
interference occurs when $Y(a_i, a_j) \neq Y(a_i, a_j')$ if $a_j \neq a_j'$.
It is a component of the well-known stable unit treatment value
assumption~\citep[SUTVA;][]{rubin1980}, commonly assumed for identification of
causal estimands, including those of
MTPs~\citep{haneuseEstimationEffectInterventions2013, young2014identification}.

Previous work has focused on settings exhibiting \textit{partial} interference
\citep{hudgensCausalInferenceInterference2008,
tchetgenCausalInferencePresence2012, halloranDependentHappeningsRecent2016},
which occurs when units can be partitioned into clusters such that interference
only occurs between units in the same cluster. Our work focuses instead on a
broader setting, that of \textit{network
interference}~\citep{vanderlaanCausalInferencePopulation2014}, which occurs
when a unit's outcome is subject to interference by other units' exposures
according to some arbitrary known network of relationships between units. When
such interference is present, the data $\Ob$ includes an adjacency matrix or
\textit{network profile}, $\Fb$, describing each unit's neighboring units,
occasionally termed
``friends''~\citep{sofryginSemiParametricEstimationInference2017}.

\cite{aronowEstimatingAverageCausal2017} demonstrate how to identify a causal
estimand when SUTVA is violated due to network interference. To do this, they
use an \textit{exposure mapping}: a function that maps the exposure assignment
vector $\Ab$ to the \textit{exposure actually received} by each unit. The
exposure received is a function of a unit's original exposure $\Ab$ and
covariates $\Lb$, including the network profile $\Fb$. If the exposure mapping
is correctly specified and consistent, then SUTVA is restored, and causal
effects subject to interference can be identified for the exposure arising
under the exposure mapping.

\cite{ogburnCausalInferenceSocial2022} and
\cite{vanderlaanCausalInferencePopulation2014} rely on similar logic to
identify population causal effects from data exhibiting a causally dependent
structure, doing so by constructing ``summary functions'' of neighboring units'
exposures. Notably, they describe semi-parametric theory
for efficiently estimating the effects of stochastic interventions in the
network dependence setting. Stochastic interventions, which differ from MTPs,
replace the natural value of exposure with a random draw from an
investigator-supplied counterfactual distribution
\citep{munozPopulationInterventionCausal2012}. While a mathematically
elegant strategy, the interpretation of a stochastic intervention is typically
challenging---at times, even impractical---as real-world policies can seldom be
defined by randomly assigning (post-intervention) exposure values to study
units. \jrssb{Furthermore, this setting gives rise to estimation challenges:
empirical evaluations in \cite{ogburnCausalInferenceSocial2022} and related work by \cite{sofryginSemiParametricEstimationInference2017} relied on relatively
bespoke parametric modeling of nuisance parameters, coupled
with Monte Carlo procedures for point and variance estimation.}

Notably, while the hypothetical exposure that results from this random
draw is not guaranteed to match that which would result from an MTP, the two
classes of interventions may be constructed to yield equivalent
counterfactual means~\citep{young2014identification}. Given the similarities
between these intervention schemes, we extend recent theoretical developments
to construct semi-parametric efficient estimators of the effects of MTPs under
network interference. Our work reveals that, for MTPs, much of the previously
established semi-parametric theory can be reduced in ways that simplify the
application of machine learning and adaptive, non-parametric regression for
nuisance estimation.

But what does it mean to intervene on a summary function? If one were
to intervene on the summary directly, the resulting collection of
counterfactual exposures could plausibly be inconsistent with the structure of
the network. Some existing works seek to circumvent this issue by
recasting the desired estimand as a mean of individual-level causal
effects~\citep{aronowEstimatingAverageCausal2017, athey2017, Savje2023}. When
investigators aim to estimate the impact of a hypothetical policy, however,
this strategy will not answer questions of scientific interest---for the
estimand does \textbf{not} correspond to a population-level intervention that
could be implemented in practice. In order to estimate a population-level
causal effect, one must consider first intervening, and only \textit{then}
applying the summary or exposure mapping---a process more naturally applicable
to MTPs.

Other relevant works address interference under different assumptions: random
networks~\citep{clarkCausalInferenceStochastic2024}, multiple
outcomes~\citep{shin2023}, long-range
dependence~\citep{tchetgentchetgenAutoGComputationCausalEffects2021,
tchetgen2025autodrestimation}, bipartite
graphs~\citep{ziglerBipartiteInterferenceAir2023}, and unknown network
structure~\citep{ohnishi2022degree, hoshino2023causal}. We build on the setting
described by \cite{ogburnCausalInferenceSocial2022}, as their scientific
goals most closely resemble those of the MTP framework.

\section{Methodology}\label{section:methods}

Suppose
there exists a network describing whether two units are causally dependent with
adjacency matrix $\Fb$, where $F_i$ denotes the neighbors of unit $i$. For each
unit $i$, a set of confounders $L_i$ is drawn, followed by an exposure $A_i$
based on a summary $L_i^s$ of its own and its neighbors' confounders, and finally
an outcome $Y_i$ based on $L_i^s$ and a summary of its own and its neighbors'
exposures, $A_i^s$. This data-generating process can be defined formally as the
SCM in Equation~\eqref{npsem2}:
\begin{equation}\label{npsem2}
    L_i = f_L(\varepsilon_{L_i});
    A_i = f_A(L_i^s, \varepsilon_{A_i});
    Y_i = f_Y(A_i^s, L_i^s, \varepsilon_{Y_i}) \ .
\end{equation}
Since $A_i^s$ and $L_i^s$ are not directly observed but are functions
that follow from the scientific problem at hand, it follows that (in a slight abuse
of notation) $A_i^s = s_{F_i}(\Ab, \Lb)$ and $L_i^s = s_{F_i}(\Lb)$, which we
use to denote that, for unit $i$, the summary function depends on $i$'s column
of the network adjacency matrix $\Fb$. This means that $A_i^s$ and $L_i^s$
depend only on neighboring units. As just one example, $A_i^s$ may be a
(possibly weighted) sum of neighbors: $s_{F_i}(\Ab, \Lb) = \sum_{j = 1}^n F_{ij}A_j =
\sum_{F_{ij} \neq 0} A_j$.

Following \cite{ogburnCausalInferenceSocial2022}, we assume error vectors
$(\varepsilon_{L_1}, \ldots, \varepsilon_{L_n})$, $(\varepsilon_{A_1}, \ldots,
\varepsilon_{A_n})$, and $(\varepsilon_{Y_1}, \ldots, \varepsilon_{Y_n})$ are
independent of each other, with entries identically distributed and either
$\varepsilon_{i} \indep \varepsilon_{j}$ provided $\{i, j\} \not\subseteq F_k,
\, \forall \,\, k \in 1, \ldots, n$ or $\text{Cov}(\varepsilon_i, \varepsilon_j) \geq 0$ otherwise. That is, errors between units are
independent provided that the units are neither directly connected nor share
ties with a common node in the interference network given by $\Fb$; otherwise, errors across units may be positively correlated. Positive correlation ensures Theorem \ref{theorem:ogburn} will hold, and is typical in applied settings with interference.

Interference bias arises when the data arise from the SCM~\eqref{npsem2}
but investigators wrongly assume that $f_Y$ is a function only
of $A_i$ and $L_i$, and not of $\{A_j \colon F_{ij} \neq 0\}$ or $\{L_j \colon
F_{ij} \neq 0\}$. Since interference violates the consistency
rule~\citep{Pearl2010}, commonly relied upon to identify causal effects,
ignoring its presence, even inadvertently, leads to a failure in
identification and consequently risks biased estimation. Under the
SCM~\eqref{npsem2}, identifiability of the causal effect of applying an
exposure to all units $A_j \colon j \in F_i$ can be restored by controlling for
all $L_j \colon F_{ij} \neq 0$ directly or via the dimension-reducing summaries
$A_i^s$ and $L_i^s$~\citep{vanderlaanCausalInferencePopulation2014} of a unit's
neighbors' confounders and exposures.

\subsection{Induced Modified Treatment Policies}\label{section:imtp}

As discussed in Section \ref{section:background}, restoring identifiability by performing inference on $A^s$ and $L^s$
instead of $A$ and $L$ is a common strategy for handling interference. This approach is, however, not
compatible with the application of MTPs: one must consider not
the causal effect of $A$ under $d(\cdot; \delta)$; but rather, the
effect of $A^s$ after intervening on the upstream exposure via $d(\cdot;
\delta)$. To identify the causal effects of MTPs under interference, we
introduce a novel intervention scheme---the \textit{induced MTP}.

Consider applying an MTP to the SCM~\eqref{npsem2}, replacing $\Ab$ with the vector $\Ab^{d} = d(\Ab, \Lb; \delta) = [d(A_i, L_i;
\delta)]_{i=1}^n$. Under interference, we are interested in the causal effect
of $A^s$ on $Y$. Hence, the scientific question of interest is actually,
``what if $A_i^s$ were replaced by some $A_i^{s \circ d} = s_{F_i}(d(\Ab, \Lb;
\delta), \Lb)$?'' This process is illustrated in Figure~\ref{fig:square}. We
call the function composition $s \circ d$ the \textit{induced MTP}.
\begin{figure}[h]
\centering
\begin{tikzpicture}
\tikzset{vertex/.style = {shape=circle,draw,minimum size=1em}}
\tikzset{edge/.style = {->,> = latex'}}
% vertices
\node[] (a) at  (0,0) {$\Ab$};
\node[] (b) at  (2,0) {$\Ab^d$};
\node[] (c) at  (4,0) {$\Ab^{s\circ d}$};

%edges
\draw[edge] (a) to (b) node[above=0.6cm, left=1cm, anchor=north] {$d$};
\draw[edge] (b) to (c) node[above = 0.6cm, left=1cm, anchor=north] {$s$};
\end{tikzpicture}
\vspace{0.1in}
\caption{How an induced MTP arises as the composition of MTP and
  summary functions $d \circ s_A$.}
\label{fig:square}
\end{figure}

\noindent
The counterfactual mean of an induced MTP is given by
Equation~\eqref{imtp}:
\begin{equation}
  \Psi_n(\Pf) = \E_\Pf \Big[\frac{1}{n}\sum_{i=1}^n
  Y(s_{F_i}(d(\Ab, \Lb; \delta), \Lb))\Big] \ .
\label{imtp}
\end{equation}
\noindent
Under an induced MTP, interference no longer hampers identifiability because
$s_{F_i}(\Ab, \Lb)$ captures the contribution of all
relevant units (that is, a given unit $i$ and its neighbors) to each $Y_i$. This
data-adaptive parameter will converge to the population counterfactual mean as
$n\rightarrow \infty$ \citep{Hubbard2016}. Use
of such a parameter definition is necessary because we must condition on the
single observation of the interference network at play. Hence, do note
that in all theory that follows, our estimates will implicitly condition on
the observed network $\Fb$. This estimand must be compatible with
the network: it is interpreted as the average change in $Y$ caused
by imposing the unit-level MTP $d$ on each unit in the population
governed by the network $\Fb$. With $\Psi_n(\Pf)$ identified, the population intervention effect (a contrast) may be
defined by subtracting $\E Y$~\citep{munozPopulationInterventionCausal2012}.

\subsection{Identification}\label{section:identification}

In addition to the SCM~\eqref{npsem2}, to identify the causal
parameter $\Psi_n(\Pf)$ by a statistical parameter $\psi_n$, we rely on the
following assumptions:%
\noindent
\assumption[Summary positivity]{
  If the summaries satisfy $(s_{F_i}(\ab, \lb), s_{F_i}(\lb)) \in \text{supp}\{A_i^s, L_i^s\}$, then the summaries after intervention satisfy
  $(s_{F_i}(\ab^d, \lb), s_{F_i}(\lb)) \in \text{supp}\{A_i^s, L_i^s\}$.%
  % \nsh{here again the notation switches from $s_A$ to $s_{F_i}$. the former,
  % with $F_i$ as an argument seems clearer to me, as using the latter for both
  % the summaries for exposure and covariates suggests that the same summary
  % function is necessary (and i think that's not the case).}
}\label{assumption:positivity}%

\noindent
Assumption~\ref{assumption:positivity} is a weaker positivity
requirement than that required for identification of the causal dose-response
curve---rather than requiring all possible exposure values to be observable
for every combination of covariates, we only require that the MTP keep the
exposure of each unit within the support defined by its \textit{own}
covariates, that is, for its own stratum. To some extent, this can be
enforced by design if the investigator were to choose $d$ appropriately.
Furthermore, Assumption~\ref{assumption:positivity} is even weaker than the
standard MTP positivity assumption: technically, $A_j^d \in \text{supp}\{A,
L\}$ for all $j \in F_i$ is not required, only that the summary $A_i^{s\circ
d}$ remains in the support of $A_i^s$. In other words, we do not require
positivity for individual exposures, \textit{only positivity on the summary as
a whole}. Importantly, $\text{supp}\{A_i^s, L_i^s\}$ denotes the support of
$A_i^s$ and $L_i^s$ together implied by the number of neighbors $F_i$, as,
under a fixed network adjacency matrix $\Fb$, the summary function that
produces $A_i^s$ may depend on $\Lb$ and $F_i$. Since
$\text{supp}\{A_i^s, L_i^s\}$ may differ for each unit depending on the number
of neighbors, this could be much smaller than $\A_i^s
\bigotimes \Li_i^s$ for certain summaries.%

\noindent
\assumption[No unmeasured confounding]{
  \jrssb{The potential outcome satisfies the condition $Y(s_{F_i}(\ab, \Lb)) \indep s_{F_i}(\Ab, \Lb) \mid \Lb$ for all  $\ab
  \in (\A)^n$ and $i \in \{1, \ldots, n\}$\ .}
}\label{assumption:confounding}

\noindent
This assumption may be interpreted as slightly weaker than the typical no
unmeasured confounding assumption in causal inference: we only require that
potential outcomes of a unit $i$ are independent of all possible exposure
summaries that could be obtained conditional on the neighbors involved in the
summary. For example, if a particular unit's summary only depends on a single
neighbor's exposure, then for that unit, the analysis need only adjust for
its neighbor's confounders in addition to its own.%

%\assumption[Locally lipschitz MTP]{\jrssb{
%The inverse of 
%$d^{-1}(\ab, \lb, \delta)$ and $s_{F_i}(\ab, \lb)$ exist and are locally Lipschitz in $\ab$, meaning that for all $\ab \in \A$, there exists a neighborhood $U \subset \A$ such that for some constant $L$,}

%\begin{equation*}
%    \rVert d(\ab_1, \lb; \delta) - d(\ab_2, \lb; \delta)\rVert < L\rVert \ab_1 - \ab_2\rVert, \forall \ab_1, \ab_2 \in U
%\end{equation*}

%}\label{assumption:psi}%

%\noindent
%\jrssb{By Rademacher's theorem, a locally Lipschitz function must be almost-everywhere differentiable. An almost-everywhere differentiable function is locally Lipschitz if its derivative is locally bounded. Alternatively, a function is locally Lipschitz if it is piecewise continuously differentiable. Hence, Assumption~\ref{assumption:psi} generalizes the ``piecewise smooth invertibility'' assumption first introduced
%by~\cite{haneuseEstimationEffectInterventions2013}, a standard assumption in the MTP
%literature~\citep{diaz2018stochastic, diazNonparametricCausalEffects2023}.}

\assumption[Piecewise smooth invertibility]{\jrssb{The MTP $d(a, l; \delta)$
    has an inverse defined
\begin{equation}
    d^{-1}(a, l; \delta) = \sum_{k=1}^{K(l)}h_k(a, l; \delta)\mathbb{I}(a \in \A_k(l)) \ ,
\end{equation}
for some countable set of functions $h_1, \ldots, h_{K(l)}$, where each $h_k$
is continuously differentiable with respect to $a$ over the partition $\A_k(l)$, a set which may depend on the covariates $l$.
}}\label{assumption:psi}%

\noindent
Assumption~\ref{assumption:psi} was first introduced
by~\cite{haneuseEstimationEffectInterventions2013} and is standard in the MTP
literature~\citep{diaz2018stochastic, hejazi2022efficient,
diazNonparametricCausalEffects2023}, where it has been used to ensure the
existence of the efficient influence function, allowing for the construction of
regular asymptotically linear estimators of the corresponding statistical
estimand.

%In Theorem~\ref{supp:theorem:absolute-continuity} of the Supplementary Material, we show  that $d$ must be absolutely continuous, and therefore differentiable almost everywhere, for $\Psi_n(\Pf)$ to be identified. By the inverse function theorem, this implies that the derivative of $d^{-1}$ must exist almost everywhere, meaning that piecewise smooth invertibility of $d$ is a necessary condition for identification.

\assumption[Summary coarea]{The \textit{coarea} of $s_{F_i}(\ab, \lb)$ exists
  and exceeds zero almost everywhere. \jrssb{That is, for almost all $\ab$,
  $s_{F_i}$ has a Jacobian $J_\ab$ with respect to $\ab$ satisfying}
\begin{equation}\label{eq:coarea}
  \sqrt{\det J_\ab s_{F_i}(\ab, \lb) J_\ab  s_{F_i}(\ab, \lb)^\top} > 0 \ .
\end{equation}
}\label{assumption:coarea}
In Equation \ref{eq:coarea}, the left-hand side is called the coarea of $s$. See~\cite{Negro2022} for a discussion of coarea in statistics.
%Theorem~\ref{supp:theorem:absolute-continuity} in the Supplementary Material~\ref{supp:section:preliminaries} proves that the existence of $J_\ab s_{F_i}(\ab, \lb)$ (and consequently, the coarea) is also a necessary condition for identification. 
Together,
Assumptions~\ref{assumption:psi} and~\ref{assumption:coarea} ensure that the
causal estimand can be expressed as an estimable function of $A_i^s$ and
$L_i^s$ instead of the entire vectors $\Ab$ and $\Lb$, and that its efficient
influence function exists, which permits the construction of regular
asymptotically linear estimators capable of achieving the semi-parametric
efficiency bound.

Under Assumptions~\ref{assumption:positivity}--\ref{assumption:coarea},
the counterfactual mean of an induced MTP $\Psi_n(\Pf)$ is identified by
\begin{equation}
    \psi_n = \frac{1}{n}\sum_{i=1}^n \E_{\Pf}(m(A_i^s, L_i^s) \cdot
      r(A_i^s, A_i^{s\circ d}, L_i^s) \cdot w(\Ab, \Lb, i))  \ ,
\end{equation}
\noindent where, letting $J_\ab$ denote the Jacobian with respect to $\ab$, the nuisances $m$, $r$, and $w$ are defined:
\begin{align}
    m(a^s, l^s) = \E_Y(Y \mid A_i^s = a^s, L_i^s = l^s) \tag{Conditional Mean}\\
    r(a^s, a^{s \circ d^{-1}}, l^s) = \frac{p(a^{s \circ d^{-1}} \mid l^s)}
      {p(a^s \mid l^s)} \tag{Density Ratio} \\
    w(\ab, \lb, i) = \sqrt{\frac{\det J_\ab (s_{F_i} \circ d^{-1})(\ab, \lb;
      \delta)J_\ab (s_{F_i} \circ d^{-1})(\ab, \lb; \delta)^\top}{\det J_\ab
      s_{F_i}(\ab, \lb)J_\ab s_{F_i}(\ab, \lb)^\top}} \tag{Induced MTP Weights}
\end{align}
See Supplementary Material~\ref{supp:section:identification} for a proof.
While $m$ and $r$ are nuisance parameters that must be estimated from
the data, $w$ is a deterministic function of the data and the choices of $s$
and $d$ made by the investigator. Since $m$ and $r$ only depend on the
summaries of $\Ab$ and $\Lb$, this identification result effectively factorizes
the estimand: \textit{all estimation can proceed using exposure and confounder
summaries} instead of their individual-level values. Furthermore, while the form
of $w$ may appear complex, if $s$ and $d$ are linear, $w$ often simplifies considerably. For example, if $d(a, l; \delta) =
\delta \cdot a$ and $s_{F_i}(\ab, \lb) = \sum_{j \in F_i} \omega_ja_j$ (a weighted sum),
then $w(\ab, \lb, i) = 1/\delta$. Unlike for the estimand
of~\cite{ogburnCausalInferenceSocial2022}, no Monte Carlo procedures are
necessary to estimate these nuisance quantities.

\subsection{Asymptotically Efficient Estimation}\label{section:efficient}

It is well-established that standard plug-in and re-weighting estimators cannot
leverage flexible machine learning or non-parametric regression strategies for
estimation of nuisance parameters without incurring possibly severe asymptotic
bias~\citep{koshevnik1977non, pfanzagl1985contributions, bickel1993efficient}.
Constructing a consistent estimator that achieves the semi-parametric
efficiency bound---the lowest possible variance among regular asymptotically
linear estimators---in the non-parametric model $\M$ requires alternative
strategies, for which the efficient influence function is a common ingredient.
Such strategies include one-step bias-corrected
estimation~\citep{pfanzagl1985contributions, bickel1993efficient}, unbiased
estimating equations~\citep{vdl2003unified}, targeted maximum likelihood (or
minimum loss) estimation~\citep{van2006targeted, van2011targeted}, and, most
recently, double machine learning~\citep{chernozhukov2018double}. In many
instances, these distinct frameworks yield doubly robust estimators.

We use \textit{doubly robust} to mean that, \jrssb{for nuisances $m$ and $r$, consistent estimation occurs if the nuisance error product $\lVert \hat{m} - m \rVert_2 \lVert \hat{r} - r \rVert_2$ is asymptotically negligible}. Many common non-parametric regression
algorithms can be shown to converge at $n^{1/4}$ rates under certain
assumptions (see, e.g., Section 4.3 of~\cite{kennedy2022semiparametric} or
Section 4.1 of~\cite{diaz2020nonparametric} and references therein).
Consequently, when such flexible algorithms are used for nuisance estimation,
the product of their convergence rates will be at least $n^{1/2}$ (or
faster), making a key second-order remainder term in the von Mises expansion of
the estimator and target parameter negligible; moreover, this allows for a
doubly robust estimator to achieve consistency under misspecification of either
of two nuisance estimators. For a doubly robust estimator to achieve the
semi-parametric efficiency bound, both nuisance estimators must be correctly
specified~\citep[see, e.g.,][]{van2011targeted, kennedy2022semiparametric}.

To construct semi-parametric efficient estimators of $\psi_n$, we build upon
theoretical developments for estimating the causal effect of a stochastic
intervention under interference. Assuming the SCM~\eqref{npsem2},
\cite{sofryginSemiParametricEstimationInference2017} showed that such a causal
effect may be identified as
\begin{equation}
  \psi_n^\star = \frac{1}{n}\sum_{i = 1}^n\int_{\Li^s}\int_{\A^s}
    m(a_i^s, l_i^s)\bar{p}^{\star}(a_i^s \mid l_i^s) p(l_i^s)
    \partial\mu(a_i^s, l_i^s) \ ,
\end{equation}
where $\bar{p}^{\star}$ is the mixture of neighbors' conditional exposure
distributions that results from replacing each $A_i$ with a random draw
$A_i^{\star}$ from some user-specified distribution. As noted by other authors~\citep{young2014identification, diaz2018stochastic},
an MTP may be expressed as a variant of a stochastic intervention whose
replacement density depends on $A^s$ and satisfies
piecewise smooth invertibility. Consequently, an induced MTP is a stochastic
intervention where%
\begin{equation}\label{eq:mtp-density}
  \bar{p}^\star(a_i^s \mid l_i^s) = p_{A^s}(a_i^{s \circ d^{-1}} \mid l_i^s)
    \sqrt{J_\ab (s_{F_i} \circ d^{-1})(\ab, \lb; \delta)J_\ab (s_{F_i}
    \circ d^{-1})(\ab, \lb; \delta)^\top} \ ,
\end{equation}%
\noindent which follows from the change-of-variables formula for
functions whose Jacobians $J_\ab$ are not square. We show this equivalence in Lemma~\ref{supp:lemma:change-of-variables} and Section~\ref{supp:section:eif} of the Supplementary Material.

The efficient influence function (EIF) for regular asymptotically linear
estimators of the statistical functional that identifies the causal effect of a
stochastic intervention under interference was first derived by
\cite{vanderlaanCausalInferencePopulation2014}. Treating the EIF as an
estimating equation is a standard strategy for the construction of
semi-parametric efficient estimators~\citep{pfanzagl1985contributions, bickel1993efficient,
van2011targeted}. Applying the representation in
\jrssb{Equation~\eqref{eq:mtp-density}}, we prove in Section~\ref{supp:section:eif}
of the Supplementary Material that the EIF for the causal effect of an
induced MTP is
\begin{equation}
  \bar{\phi}_\Pf = \frac{1}{n}\sum_{i=1}^n
    r(A_i^s, A_i^{s\circ d^{-1}}, L_i^s)w(\Ab, \Lb, i)(Y_i - m(A_i^s, L_i^s)) +
    \E_{\Pf}[m(A_i^{s \circ d}, L_i^s) \mid \Lb = \lb] - \psi_n \ .
\label{eq:eif}
\end{equation}%
\noindent
The EIF $\bar{\phi}_\Pf$ is expressed as an empirical mean because it represents
the influence of all $n$ units within the \textit{single draw} of the
interference network that is observed. Despite only observing a single
realization of the network, it is clear that this EIF is composed of individual
$i$-specific components, with the corresponding EIF estimating equation
admitting the expression:
\begin{equation}
  \bar{\phi}_\Pf = \frac{1}{n}\sum_{i=1}^n \phi_\Pf(O_i) - \psi_n = 0 \ ,
\label{eq:eif-equation}
\end{equation}%
where $\phi_\Pf(O_i) = r(A_i^s, A_i^{s\circ d^{-1}}, L_i^s)w(\Ab, \Lb,
i)(Y_i - m(A_i^s, L_i^s)) + \E_{\Pf}[m(A_i^{s\circ d}, L_i^s) \mid \Lb =
\lb]$. \cite{ogburnCausalInferenceSocial2022} exploit this structure to prove the following theorem, reproduced for convenience:
\begin{theorem}[Central Limit Theorem of~\cite{ogburnCausalInferenceSocial2022}]
  Suppose an estimator $\hat{\psi}_n$ is a solution to the EIF estimating
  equation of Equation~\eqref{eq:eif-equation} and that
  $K_{\text{max}}^2 / n \rightarrow 0$ as $n \rightarrow \infty$, where
  $K_{\text{max}}$ denotes the maximum node degree in the network. Under the
  SCM~\eqref{npsem2} and mild regularity conditions, including that the estimating function is bounded, 
  \begin{equation}
    \sqrt{C_n}(\hat{\psi}_n - \psi_n^\star) \overset{d}{\rightarrow}
    N(0, \sigma^2) \ ,
  \end{equation}
  
  \noindent for some finite $\sigma^2$ and some constant $C_n$ such that
  $n/K_{\text{max}}^2 \leq C_n \leq n$ \ .
\label{theorem:ogburn}
\end{theorem}%

\noindent
This theorem states that if $K_{\text{max}}$ grows asymptotically no faster
than $n^{1/2}$, then an estimator $\hat{\psi}_n$ of $\psi_n^\star$ attains a
normal limiting distribution centered about $\psi_n^\star$. In addition, these
authors argue, based on earlier results
of~\cite{vanderlaanCausalInferencePopulation2014}, that such an estimator of
$\psi_n^\star$ will exhibit double robustness with respect to the nuisance
estimators $\hat{m}$ and $\hat{r}$, with a second-order remainder term
of the form $\lVert \hat{m} - m \rVert_2 \lVert \hat{r} - r \rVert_2 =
o_{\Pf}(\sqrt{C_n})$, where the estimator $\hat{\psi}_n$ of $\psi_n^\star$
remains consistent as long as the rate-product of the differences of the
nuisance estimators $\{\hat{m}, \hat{r}\}$ from their respective targets
$\{m, r\}$ converges at the specified rate. Since our target estimand $\psi_n$
is equivalent to $\psi^\star_n$ under an MTP and appropriate identification
conditions, using Theorem~\ref{theorem:ogburn} and the EIF in
Equation~\eqref{eq:eif}, we can construct semi-parametric efficient estimators
of $\psi_n$ in at least two ways, which we outline next.

\textbf{One-Step Estimation}. A one-step bias-corrected
estimator~\citep{pfanzagl1985contributions, bickel1993efficient} uses
Equation~\eqref{eq:eif-equation} to de-bias an initial plug-in estimator by adding to it an estimated $\phi_{\hat{\Pf}_n}(O_i)$, constructed
based on nuisance estimators. While we use $\Pf_n$ to denote the empirical
distribution, we use $\hat{\Pf}_n$ to denote the empirical distribution
augmented by relevant nuisance estimators. The outer expectation in the term
$\E_{\Pf}[\hat{m}(A_i^{s\circ d}, L_i^s) \mid \Lb = \lb]$ in the EIF can
be dropped because the sample means of $\hat{m}(A_i^{s \circ d}, L_i^s)$ and
$\E_{\Pf}[\hat{m}(A_i^{s\circ d}, L_i^s) \mid \Lb = \lb]$ both converge to the
same value (see Section~\ref{supp:section:one-step} of the Supplementary
Material for a proof). Then, the one-step (OS) estimator is
\begin{equation}
  \hat{\psi}_n^{\text{OS}} = \frac{1}{n}\sum_{i=1}^n \phi_{\hat{\Pf}_n}(O_i) \ ,
  \label{eq:one-step}
\end{equation}
where $\phi_{\hat{\Pf}_n}(O_i) = \hat{r}(A_i^s, A_i^{s\circ d^{-1}}, L_i^s)w(\Ab, \Lb, i)(Y_i -
\hat{m}(A_i^s, L_i^s)) + \hat{m}(A_i^{s\circ d}, L_i^s)$. Since the estimating functions are no longer centered,
variance estimates must be adjusted (as described in
Section~\ref{section:variance}).

\textbf{Targeted Maximum Likelihood Estimation (TMLE)}. Although the one-step
estimator is semi-parametric efficient, it is not a substitution estimator: it
may yield estimates outside the bounds of the parameter space.
TMLE---a general template for the construction of substitution estimators that
appropriately solve the EIF estimating equation~\citep{van2006targeted,
van2011targeted}---resolves this shortcoming. Estimation under an induced MTP
proceeds as follows:
\begin{enumerate}
  \item Estimate nuisances functions to obtain $\hat{r}(A_i^s, L_i^s)$, $\hat{r}(A_i^{s\circ d},
    L_i^s)$
    $\hat{m}(A_i^s, L_i^s)$, and $\hat{m}(A_i^{s\circ d}, L_i^s)$, and
    compute weights $w(\Ab, \Lb, i)$ and $w(\Ab^d, \Lb, i)$ from the data based on the chosen $s$ and $d$. 
  \item Fit a one-dimensional parametric fluctuation model regressing
    $\hat{r}(A_i^s, L_i^s)w(\Ab, \Lb, i)$ on $Y_i$ with offset $\hat{m}(A_i^s,
    L_i^s)$; a common choice for this is logistic regression
    $\logit(Y_i) = \logit(\hat{m}(A_i^s, L_i^s)) + \varepsilon \hat{r}(A_i^s,
    L_i^s)w(\Ab, \Lb, i)$, where the $Y_i$ are rescaled to ensure each lies in
    the open unit interval $(0,1)$.
  \item Compute $\hat{m}^{\star}(A_i^{s\circ d}, L_i^s)
    = \text{expit} (\logit(\hat{m}(A_i^{s\circ d}, L_i^s) +
    \hat{\varepsilon} \hat{r}(A_i^{s\circ d},
    L_i^s)w(\Ab^d, \Lb, i)$ \jrssb{if using the logistic fluctuation model, or more broadly, compute the corresponding parametric fluctuation model predictions under the MTP intervention}, based on the MLE
    $\hat{\varepsilon}$ of $\varepsilon$.
  \item Compute the TML estimate as $\psi_n^{\text{TMLE}} =
    \frac{1}{n}\sum_{i=1}^n \hat{m}^{\star}(A_i^{s\circ d}, L_i^s)$.
\end{enumerate}%
\noindent
In Step 2, the corresponding estimate $\hat{\varepsilon}$ is used to fluctuate
the initial estimator $\hat{m}(A_i^s, L_i^s)$, using the summary-weighted
density ratio $\hat{r}(A_i^s, L_i^s)w(\Ab, \Lb, i)$, to an updated version
$\hat{m}^{\star}(A_i^s, L_i^s)$ in such a way that the EIF estimating equation
is solved. Logistic regression is commonly employed to guarantee each
prediction $\hat{m}^\star$ falls in the bounds of the parameter space, thus
respecting global constraints~\citep{Gruber2010}. One can also fit an
intercept-only model with $\hat{r}(A_i^s, L_i^s)w(\Ab, \Lb, i)$ as weights
themselves, which may improve stability. In what follows, we refer to this
estimator as ``network-TMLE,'' following nomenclature introduced by
\cite{zivichTargetedMaximumLikelihood2022}, who discuss its use in practical
settings and examine its properties in simulation experiments.

\subsection{Nuisance Parameter Estimation}\label{section:nuisances}

The underlying form of nuisance parameters $m$ and $r$ is usually recognized as
being unknown and may involve, for example, complex nonlinear interactions
between covariates. Therefore, it is desirable to estimate such quantities
using flexible regression or machine learning approaches. This process often
employs $K$-fold cross-fitting~\citep{zheng2010asymptotic,
chernozhukov2018double} to \jrssb{eliminate asymptotic bias arising from
empirical process terms}~\citep{pfanzagl1985contributions,
klaassen1987consistent}.

\jrssb{Perhaps surprisingly, the form of cross-fitting used for independent
units still accomplishes the same goals even in our network setting. That is,
splitting the data into $K$ folds, fitting nuisance functions on each
combination of $K-1$ folds, and obtaining out-of-sample predictions for each
held-out fold is sufficient to render any empirical process bias asymptotically
negligible, even when units may be correlated according to
Equation~\ref{npsem2}. A formal proof can be found in
Section~\ref{supp:section:crossfitting} of the Supplementary Material. At a
high level, it follows similarly to an $m$-dependence argument: a sample mean
is unbiased even under dependence, and its variance will shrink at a rate
dependent on the maximum degree of the network, which in our setting is limited
by assumption. The result mirrors similar numerical findings in the longitudinal data
setting~\citep[e.g.,][]{Fuhr2024}. Consequently, cross-fitting can be performed
identically whether units may be independent or correlated according to a network whose maximum degree grows according to Theorem \ref{theorem:ogburn}.}

Since the outcome regression $m$ is a conditional expectation function, it can
be estimated using any supervised learning algorithm. We recommend super
learning~\citep{vanderlaanSuperLearner2007}, that is, fitting an ensemble of
candidate regression algorithms in a cross-validated manner and selecting the
candidate yielding the lowest cross-validated risk (for continuous outcomes, we
use mean-squared error). This guarantees that the selected algorithm performs
asymptotically as well as the best candidate algorithm in the library used to
construct the ensemble~\citep{vanderLaan2004,
vaartOracleInequalitiesMultifold2006}, even in the fixed regression design
setting presently considered~\citep{daviesOptimalSpatialPrediction2016}.
\cite{phillipsPracticalConsiderationsSpecifying2023a} proposed guidance and
heuristics to aid in overcoming the considerable challenge of assembling a
candidate library from the diversity of learning algorithms available.

An advantage of the MTP framework is that the implied form of the intervention,
unlike general stochastic
interventions~\citep{munozPopulationInterventionCausal2012,
sofryginSemiParametricEstimationInference2017,
ogburnCausalInferenceSocial2022}, facilitates direct estimation of the density
ratio $r$, circumventing the need to learn a conditional density function. A
common density ratio estimator is based on probabilistic
classification~\citep{qin1998, Cheng2004}, in which a classifier is trained to
distinguish between natural and intervened samples, and its output is
transformed into a density ratio using Bayes' rule. The use of this method with
an MTP is described in detail by~\citet[][see Section
5.4]{diazNonparametricCausalEffects2023}. We recommend super learning with
binary log loss to select the optimal classifier. One can also employ
kernel-based methods, including kernel mean matching, Kullback-Leibler
importance estimation, and least-squares importance fitting, as outlined in
detail by~\cite{Sugiyama2012}.

\subsection{Variance Estimation}\label{section:variance}

Summary measures are correlated if they aggregate the same neighboring units. As a result,
the as-iid sample variance of the estimated EIF will be conservative,
especially if the network degree distribution is highly
skewed~\citep{sofryginSemiParametricEstimationInference2017}. A consistent
variance estimator for the solution to an arbitrary estimating equation
$\frac{1}{n}\sum_{i=1}^n \varphi_i = 0$ with a centered estimating function
$\varphi_i$ satisfying $\E[\varphi_i] = 0$ is given by
\begin{equation}
  \hat{\sigma}^2 = \frac{1}{n^2}\sum_{i,j} G(i, j)\varphi_i\varphi_j \ ,
\label{eq:var-est}
\end{equation}
where $G(i, j) = 1$ if $i = j$, $i$ and $j$ are in each others' dependency
neighborhoods, or $i$ and $j$ share a friend $k$, and $G(i, j) = 0$ otherwise.
A useful feature of $\hat{\sigma}^2$ is that it
automatically incorporates the scaling factor $C_n$ based on the CLT of
\cite{ogburnCausalInferenceSocial2022}. Intuitively, this is because
$\hat{\sigma}^2$ includes $n$ ``variance'' terms $\varphi_i^2$, as well as a
certain number of ``covariance'' terms $\varphi_i \varphi_j$ (for $i
\neq j$), the number of which scales with $n/K_{\max}^2 \leq C_n \leq n$,
representing the rate of connectivity in the network. Consequently,
$(1-\alpha)$\% Wald-style confidence intervals can be constructed via the
standard approach: $\hat{\psi}_n \pm
\Phi^{-1}(\alpha/2)\sqrt{\hat{\sigma}^2/n}$, where $\Phi$ is the CDF of the
standard normal distribution.

As the estimators from Section~\ref{section:efficient}
are constructed based on
the \textit{uncentered} estimating function $\phi_\Pf(O_i)$, this result does
not transport directly to our setting. However, examining the
SCM~\eqref{npsem2} reveals that $O_i$ and $O_j$ will be
identically distributed and therefore have the same mean provided that they
have the same number of neighbors, that is, $\lvert F_i \rvert = \lvert F_j
\rvert$. Following a similar strategy as~\cite{emmenegger2025}, a centered
estimating function is $\phi_{\Pf}(O_i) - \psi_n(\lvert F(O_i) \rvert)$, where
$\psi_n(\lvert F(O_i) \rvert)$ denotes the mean of $\phi_\Pf(O_j)$ over all
units in $j \in 1, \ldots, n$, satisfying the equality $\lvert F_j \rvert =
\lvert F_i \rvert$. This centering term can be estimated by computing
$\hat{\psi}_n^{\text{OS}}$ within each subgroup of possible $\lvert F_i \rvert$.
Formally, this is
\begin{equation}
  \hat{\psi}_n(\lvert F_i \rvert) = \frac{1}{|\mathcal{N}(|F_i|)|} \sum_{j \in
  \mathcal{N}(\lvert F_i \rvert)} \phi_{\hat{\Pf}_n}(O_j) \ ,
\end{equation}
where $\mathcal{N}(m) = \{k : k \in 1, \ldots, n, |F_k| = m\}$. Then, letting
$\varphi_i = \phi_{\hat{\Pf}_n}(O_i) - \hat{\psi}_n(|F_i|)$, the
estimator $\hat{\sigma}^2$ in Equation~\eqref{eq:var-est}
consistently estimates the variances of $\hat{\sigma}^\text{OS}_n$ and
$\hat{\sigma}^\text{TMLE}_n$, provided both nuisance estimators are consistent
for their targets (see Supplementary Material,
Section~\ref{supp:section:variance-estimator}, for a proof).

\section{Results: Numerical Experiments}\label{section:sim-results}

We now empirically evaluate the estimators described in
Section~\ref{section:methods}. All data in this
section are simulated in the numerical computing language
\texttt{Julia}~\citep{Bezanson2017}, using the package
\texttt{CausalTables.jl}~\citep{Balkus2025}. MTP estimates are computed using
\texttt{ModifiedTreatment.jl}~\citep{Balkus2024b}, a \texttt{Julia} package
implementing the estimators described in this work. Unless otherwise specified,
conditional means and density ratios are estimated using
super learning---from an ensemble consisting of a GLM, random forest, and multiple
gradient boosted tree models with various sets of hyperparameters---to select the algorithm
minimizing the cross-validated empirical risk with respect to an appropriate
loss (MSE for the conditional mean, binary log-loss for the density
ratio). Code for these experiments and the data analysis
is available on GitHub at
\url{https://github.com/salbalkus/pub-code-mtp-interfere}.

\subsection{Results of Experiments with Synthetic Data}\label{section:synthetic-data}

First, we evaluate estimators on simulated data with both network interference
and nonlinear relationships between confounders, exposure, and outcome. We
simulate this data using three common network structures: Erd\H{o}s-R\'enyi
(with $p = 3/n$), static scale-free (with $\lambda = 3.5)$, and Watts-Strogatz
(with $K = 6$ and $\beta = 0.5$). Data are generated according to the following
set of structural equations:

\begin{gather*}
\begin{aligned}
    \Lb_1 &\sim\text{Beta}(3,2);
    \Lb_2 \sim \text{Poisson}(100);
    \Lb_3 \sim \text{Gamma}(2,4);
    \Lb_4 \sim \text{Bernoulli}(0.6) \\
    m_L &=\Big(1 + L_4\Big)\cdot\Big(-2(\I(L_1 > 0.3) + \I(L_2 > 90) + \I(L_3>5))  -(\I(L_1 > 0.5) + \\&\I(L_2 > 100) + \I(L_3 > 10)) + 2(\I(L_1 > 0.7) + \I(L_2 > 110) + \I(L_3 > 15))\Big)\\
    \Ab &\sim \text{Normal}(m_L - 5, 1.0) \,\, \text{and} \,\,
    \Ab^s = \Big[\sum_{j \in F_i} A_i\Big]_{i = 1}^n \\
    m_A &= -2\I(A > -2) - \I(A > 1) + 3\I(A > 3); m_{A_s} = 3\I(A_s > 0) + \I(A_s > 6) + \I(A_s > 12) \\
    \Yb &\sim \text{TruncNormal}(m_L\cdot(1 + 0.2m_A + m_{A_s}) + 5, 2.0) \ ,
\end{aligned}
\end{gather*}

\noindent where $\text{TruncNormal}$ denotes a normal distribution truncated at six standard deviations. We estimate the counterfactual mean under the MTP
$d(a, l) = a + 0.25$ using network-TMLE. Being very similar, one-step estimator results are omitted for brevity.

\begin{figure}[htp]
\centering
\includegraphics[scale=0.9]{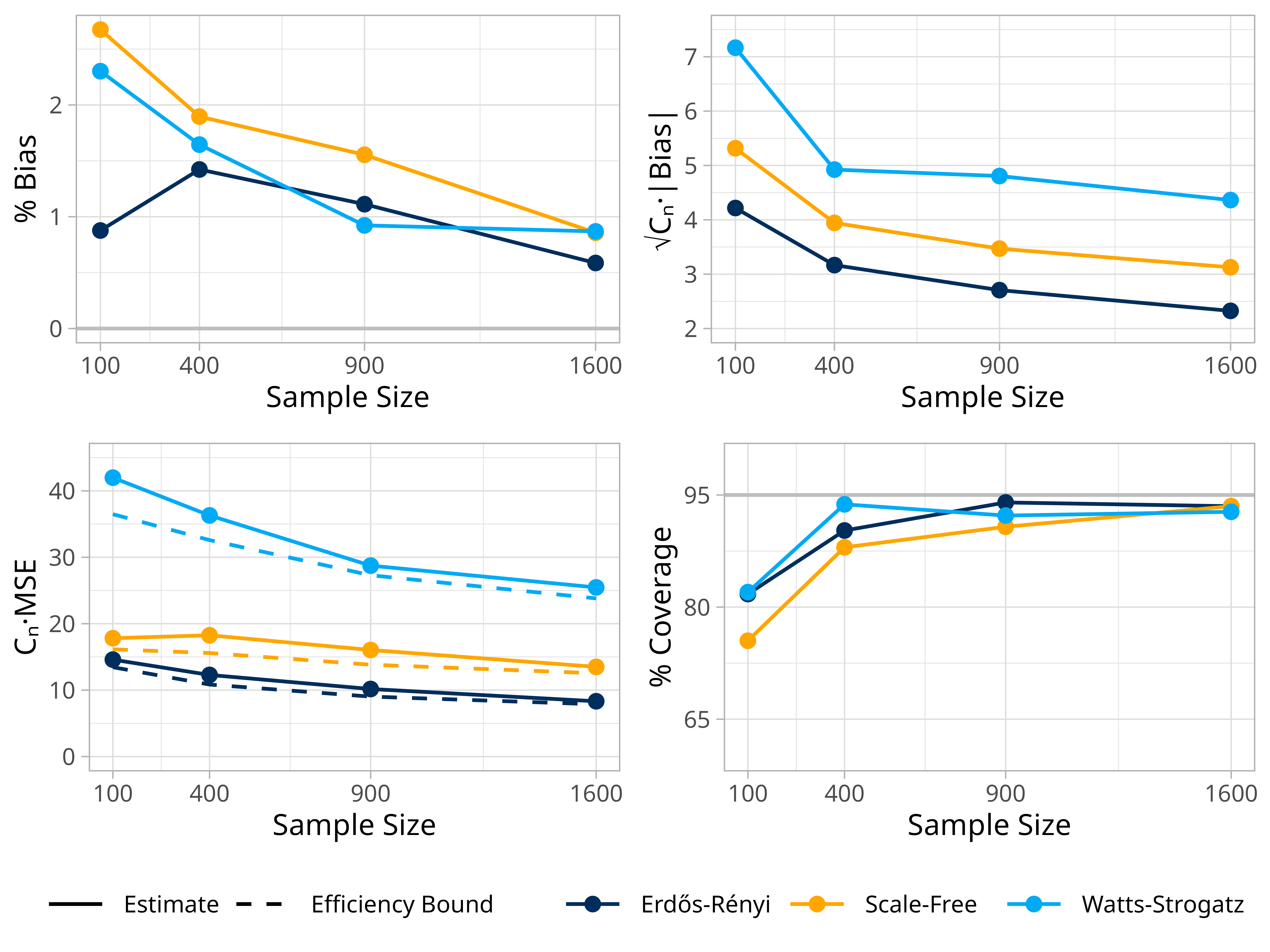}
\caption{Asymptotic performance of network-TMLE in simulation on multiple network structures.}
\label{fig:synthetic}
\end{figure}

Figure~\ref{fig:synthetic}
demonstrates how the empirical performance of our estimator using network-TMLE
aligns with theoretical results outlined in
Section~\ref{section:methods}. Bias reliably
converges towards zero with increasing sample size across all three network
structures, and is highest for scale-free, the network with the most skewed distribution of node degree. Since $K_{\max}$ in each network grows at
roughly $\log(n)$ or slower, we use $\sqrt{C}_n = \sqrt{n} / \log(n)$ as a scaling factor for the scaled bias and MSE; these values still decrease, so we can conclude that the convergence rate of network-TMLE matches the expected rate. MSE converges to the efficiency bound; see Supplementary Material \ref{supp:section:eff-bound-elaboration} for details on this bound. The MSE is highest in the Watts-Strogatz simulation, whose network has the highest average node degree (and thus the greatest correlation between units)---hence, a denser network increases variance. Furthermore, the
coverage rate of the 95\% confidence interval approaches the nominal level for all three graphs.
Based on these observations, we conclude that the network-TMLE estimator performs as
expected (see Supplementary Material \ref{supp:section:more-sim-results} for a comparison to classical methods). 

\subsection{Results of Experiments with Semi-Synthetic Data}
In a second simulation experiment, we generate semi-synthetic data based on the
real-world dataset that inspired this work and that is analyzed in
Section~\ref{section:data-analysis}. Sixteen normalized confounding variables are taken as
fixed covariates, with the 2013 and 2019 LODES commuting pattern
data~\citep{lodes_data} for California as the network profile $\mathbf{F}$. From these, we
draw exposure and outcome from normal distributions with means following a
linear function of the confounders and $A^s$, a trimmed sum of neighbors'
values of $A$ over neighbors that contributed at least \>2.5\% of commuters into
the given unit. These are represented by structural equations, conditional on
$\Lb$ and $\Fb$, expressed as

\begin{gather}
\begin{aligned}
    \Lb_k^s &= \Big[\sum_{j \in F_i} L_{k,i} \Big]_{i=1}^n \,\,\Ab \sim \text{Normal}\Big( \sum_{k=1}^{16}\Lb_k - 50 , 1.0\Big) \,\, \text{and} \,\,
    \Ab^s = \Big[\sum_{j \in F_i} A_i\Big]_{i = 1}^n \\
    \Yb &\sim \text{TruncNormal}\Big(\Ab + \Ab^s + \sum_{k=1}^{16}\Lb_k + \sum_{k=1}^{16}\Lb_k^s- 50  , 1.0\Big) \ ,
\end{aligned}
\end{gather}

We estimate the population causal effect of an MTP in 400 simulations using
four different estimators: network-TMLE with correctly-specified GLMs (linear
regression for the outcome regression and logistic regression for the density
ratio probabilistic classifier), network-TMLE with super learning for both
nuisance estimators, IID-TMLE with correctly-specified GLMs, and main-terms
linear regression. In this simulation, $n = 1652$, the same as in the dataset
considered in Section~\ref{section:data-analysis}.

Table~\ref{table:semisynthetic} depicts the operating characteristics
of the estimators. It is clear from inspection that accounting for
network interference can yield dramatically lower bias, even when the outcome
regression is otherwise correctly specified. In this case, using network-TMLE
with a correctly specified GLM decreased the mean percent bias from over 20\%
to about 0.1\%, and cut variance by about 25\%. Even if
super learning is used, the mean percent bias of the network-TMLE remains
relatively low, at around 1\%, and the variance is approximately identical.
Furthermore, confidence interval coverage for the network-TMLE is close to
nominal under both configurations of nuisance estimators. Alternative methods
suffer greatly reduced coverage due to their severe bias.

\begin{table}[htp] \centering 
  \caption{Network-TMLE performance on semi-synthetic data versus competing
  estimators}
  \label{table:semisynthetic}
\begin{tabular}{@{\extracolsep{5pt}} ccccccc}
\\[-1.8ex]\hline
\hline \\[-1.8ex]
 & Method & Learner & Bias (\%) & Variance & Coverage (\%) & CI Width \\
\hline \\[-1.8ex]
& \multirow{2}{*}{Network TMLE} & Correct GLM & $0.11$ & $1.56$ & $96.2$ & $4.88$ \\
&  & Super Learner & $1.03$ & $1.56$ & $94.0$ & $4.88$ \\
& Classical TMLE & \multirow{2}{*}{Correct GLM} & $20.43$ & $2.11$ & $54.8$ & $5.70$ \\
& Linear Regression & & $20.62$ & $2.12$ & $55.0$ & $5.71$ \\
\hline \\[-1.8ex]
\end{tabular}
\end{table}

\section{Illustrative Data Analysis: Mobile Source Air
  Pollution}\label{section:data-analysis}

In this section, we use induced MTPs to analyze the causal effect of
``zero-emissions'' vehicle (ZEV) uptake in California on NO\textsubscript{2}
air pollution using observational data. NO\textsubscript{2} is a byproduct of
gasoline-powered vehicles shown to be associated with adverse health outcomes,
including respiratory issues and mortality~\citep{Hesterberg2009,
GillespieBennett2010, Faustini2014}. In the U.S., national ambient air quality
standards (NAAQs) are set by the Environmental Protection Agency (EPA) to limit
the amount of allowable NO\textsubscript{2} in the air~\citep{NAAQS}.

\subsection{Applied Science Background and Motivations}

As ZEVs do not produce tailpipe emissions, governments have heavily promoted
their adoption to limit air pollution. State governments must ascertain whether
these policies work as intended; hence, understanding how much localized air
pollution has decreased as a result of increased ZEV adoption is of both
scientific and policy interest. \cite{garciaCaliforniaEarlyTransition2023}
analyzed whether ZEV uptake was associated with NO\textsubscript{2} air
pollution across ZIP code tabulation areas (ZCTA) in California from 2013 to
2019. They found that an increase of 20 ZEVs per 1000 population units was
associated with a reduction of 0.41 parts per billion in NO\textsubscript{2}
within a given ZCTA on average; however, with a $p$-value of 0.252 and
confidence interval of $(-1.12, 0.29)$, this finding was far from statistically
significant.

This is a problem in which network interference arises as a result of people
driving outside the ZCTA in which they reside; that is, individuals commute in
their vehicles to other ZCTAs, emitting pollution there as well.
\cite{garciaCaliforniaEarlyTransition2023} controlled for confounding using
linear regression, without accounting for interference due to individuals
commuting in vehicles from one ZCTA unit to another. As evidenced by our
simulation experiments in
Section~\ref{section:sim-results},
neglecting interference can induce bias and yield misleading confidence
intervals. In addition, overreliance on restrictive parametric modeling may be
cause for concern, as linearity imposes a strict functional form which may not
adequately capture the underlying complexity, resulting in model
misspecification bias. The goal of our data analysis is to study the same
research question as~\cite{garciaCaliforniaEarlyTransition2023} and compare how
flexibly evaluating the causal effect of an induced MTP can improve inference
in this setting.

\subsection{Estimation Strategy}

Following~\cite{garciaCaliforniaEarlyTransition2023}, we compute the exposure,
the percentage of light-duty vehicles in California actively registered as ZEVs
by April 2019, for each ZCTA from the~\cite{zev_data}. The outcome, change in
NO\textsubscript{2} from 2013 to 2019 in parts per billion (ppb), is spatially
aggregated from a 1 km grid of estimated pollution levels~\citep{no2_2013,
no2_2019}; this provides finer-grain ZCTA-level estimates than the raw sensor
data used by~\cite{garciaCaliforniaEarlyTransition2023}. Socioeconomic
confounders are accessed from the U.S.~Census via~\cite{census_data}, while
land-use confounders are obtained from the~\cite{smartlocation_data}; see
Section~\ref{supp:section:data-sources} of the Supplementary Material for
summary statistics. We performed spatial alignment and areal weighted
interpolation of missing values using the \texttt{sf}~\citep{sf} and
\texttt{areal}~\citep{areal} packages in the \texttt{R} language for
statistical computing and graphics~\citep{r}.

Finally, to account for interference, we rely on commuting networks describing
the number of people traveling between home and work ZCTAs~\citep{deSouza2023,
lodes_data}. The induced MTP summary estimates the percentage of incoming work
commuters driving ZEVs during the 2013--2019 time period, as used for the
outcome. This was computed by summing the percentage of ZEVs in each
neighboring ZCTA in each year's network, with each unit's summand weighted by
the percentage of its neighbor's population who drive to work and normalized.
Sums of neighboring covariates are also included as potential confounders for
the induced MTP.

We compared the following data-analytic strategies: (1) estimating an induced
MTP using network-TMLE, (2) estimating a classical MTP using IID-TMLE, and (3)
estimating a main-terms GLM regression coefficient. Each seeks to answer the
same question: ``how much \textit{more} would average NO\textsubscript{2} have
decreased from 2013 to 2019 had each ZCTA experienced an increase in
the percentage of vehicles registered as ZEV by 2019?'' Strategy
(1) accounts for interference and estimates nuisances using flexible super
learning. Strategy (2) uses super learning but ignores interference. Strategy
(3) imposes a strict linearity assumption and ignores interference. Both super learning procedures selected random forest for the outcome regression and density ratio probabilistic classifier.

\subsection{Analytic Results}

\begin{figure}[h]
\centering
\includegraphics[scale=0.9]{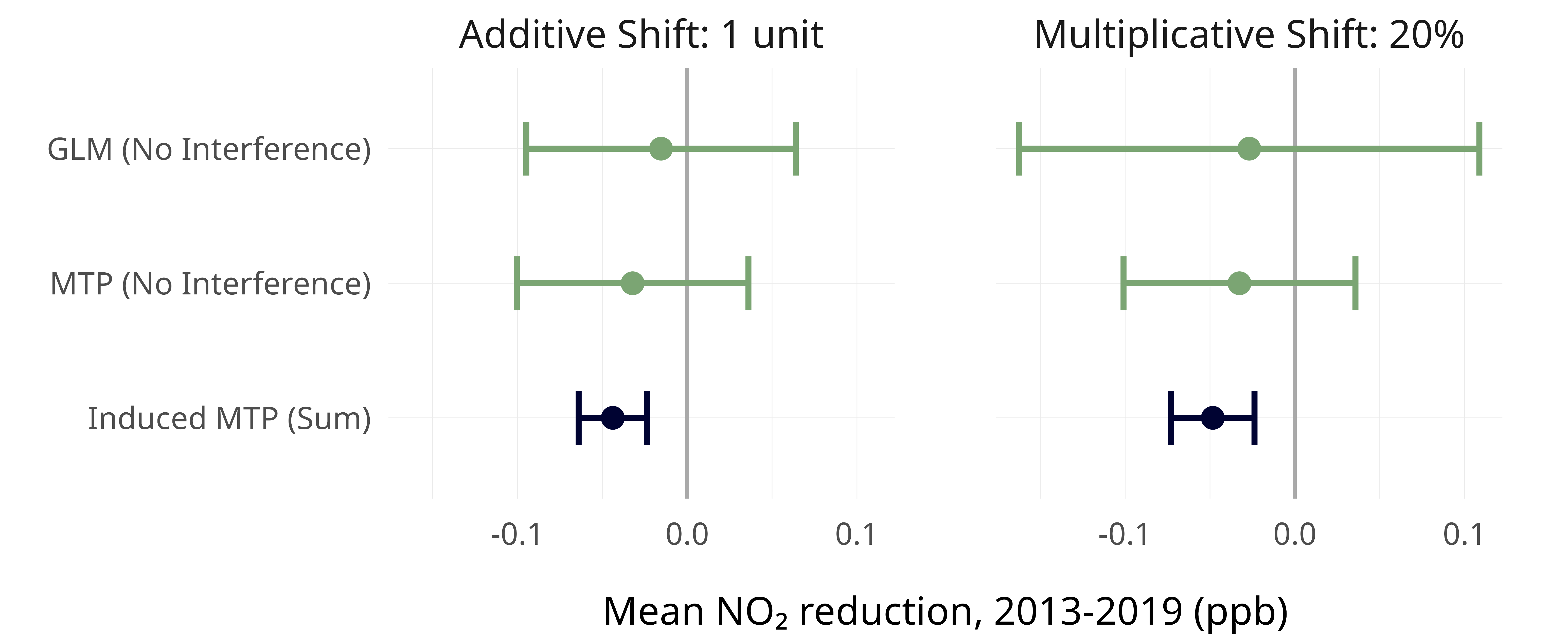}
\caption{Estimated effect sizes measuring the expected difference in
  NO\textsubscript{2} across California ZCTAs caused by two different increases
  in the proportion of ZEVs in 2019.}
\label{fig:MTPcompare}
\end{figure}

Figure~\ref{fig:MTPcompare} displays the three estimates of additive and
multiplicative effects of an MTP increasing the proportion of ZEVs. According
to the induced MTP effect estimate, adding 1\% to the proportion of ZEVs across
ZCTAs would be expected to yield a 0.044 ppb decrease in NO\textsubscript{2} on
average. This estimate is over 1.3 times larger than the MTP effect estimate
arrived at when not accounting for interference (0.032 ppb), and about 2.8
times larger than the GLM-based estimate (0.015 ppb). Similarly, the induced
MTP effect estimate indicated that scaling the proportion of ZEVs by 20\%
across ZCTAs would be expected to yield a 0.048 ppb decrease in
NO\textsubscript{2} on average---over 1.4 times larger than the naive MTP
effect estimate (0.033 ppb) and the GLM-based estimate (0.027 ppb).
Interpreted in a scientific context, for the additive shift, the estimated
effect based on the induced MTP indicates that ZEV uptake would have
accounted for about 7\% of the overall mean decrease in NO\textsubscript{2} from
2013-2019, whereas a GLM-based estimate would have accounted for only 2.5\%.

\begin{figure}[h]
\centering
\includegraphics[scale=0.9]{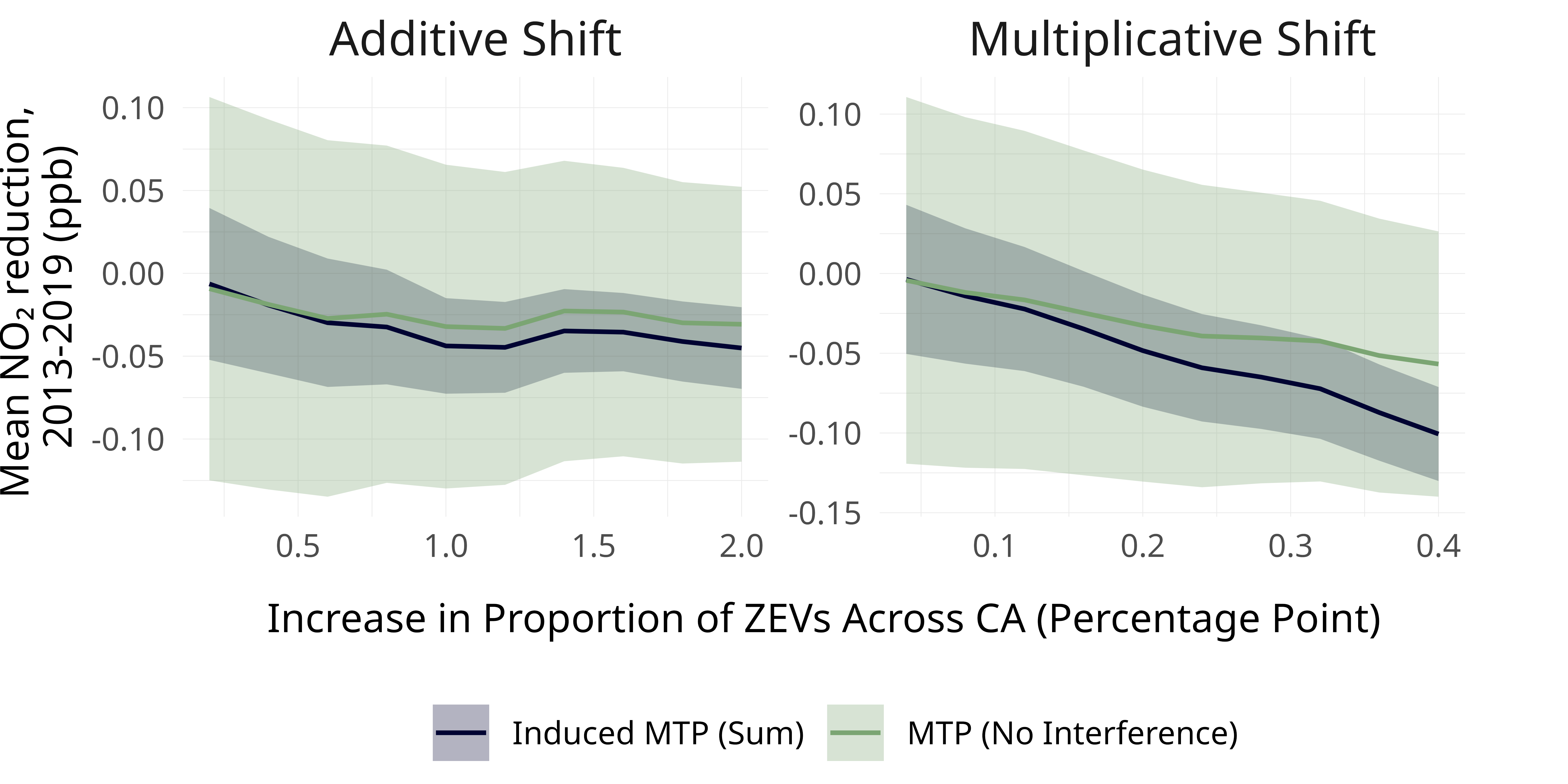}
\caption{Estimated effect sizes measuring the expected difference in NO\textsubscript{2} across California ZCTAs caused by various
  additive or multiplicative shifts in the percentage of ZEVs in 2019.
  Confidence bands use a conservative Bonferroni multiplicity
  correction.}
\label{fig:MTPgrid}
\end{figure}
Accounting for interference and using flexible regression
yielded larger effect estimates than those recovered by classical
analyses, suggesting commuting contributes significantly to vehicular NO\textsubscript{2} pollution. This was especially pronounced for the multiplicative shift,
indicating that ZCTAs with many out-commuters exacerbate NO\textsubscript{2}. Furthermore, while one might expect that accounting for interference
would yield wider confidence intervals due to correlation between units,
we observe the opposite---drastically lower variance. We conjecture that improvements in estimation of the outcome regression drive this result, as including exposure summaries as covariates reduces model misspecification, allowing the model to explain a larger proportion of the outcome variance. 

Figure~\ref{fig:MTPgrid} displays effect estimates over a grid of possible
additive and multiplicative shifts. At low-magnitude shifts, the estimates were
roughly the same. However, at larger shifts---about 0.75\% on the additive
scale and 10\% on the multiplicative scale---the estimated effects diverged,
with those accounting for interference becoming more pronounced. The trend was also less erratic under the multiplicative shift, possibly because it only produces a large shift among ZCTAs
with a large proportion of ZEVs, thereby avoiding destabilizing empirical positivity issues. Such a result highlights the importance of
considering MTPs that can be easily estimated from the data
available.

Our analysis focused on NO\textsubscript{2} since it is a well-understood pollutant. That said, from 2013--2019, average
NO\textsubscript{2} levels changed little across ZCTAs, as all ZCTAs within
California have remained well within limits regarded as safe based on EPA
guidance~\citep{NAAQS}, so the effects are inherently small. However, suppose policymakers sought further
reductions and would impose a ZEV-promoting policy given statistically significant evidence. Then, according to
Figures~\ref{fig:MTPcompare} and~\ref{fig:MTPgrid}, evidence from an induced MTP would have resulted in imposing a ZEV-promoting policy ($p <
0.05$) versus not imposing such a policy ($p > 0.05$) if interference was
ignored. Hence, our method provides much stronger and more conclusive evidence regarding even the small effect sizes in this setting. 

\section{Discussion}\label{section:discussion}

In this work, we introduced the induced modified treatment policy, a new class
of MTP that accounts for known network interference. This intervention is
useful for causal inference in observational data settings that feature
continuous exposures and network interference. We established
identification of the causal effect of an induced MTP using a novel
application of the coarea formula and outlined procedures for constructing
semi-parametric-efficient estimators capable of incorporating
flexible nuisance estimation strategies via, for example, machine learning or
non-parametric regression.

Using simulation experiments, we showed that interference can result in
significant bias when it is not corrected using an induced MTP. Our
illustrative data analysis demonstrates the perils of ignoring interference and
applying restrictive parametric modeling strategies with observational spatial
data, as is often done in environmental epidemiology and related fields. Such
oversimplifications can suggest starkly different scientific conclusions than
those provided by our proposed strategy.

In practice, several challenges remain. One limitation is that ratios of
conditional densities are still more difficult to estimate than conditional
expectation functions, especially in high-dimensional
settings~\citep{Sugiyama2012}. This problem is exacerbated under
interference: for units with an excessively large number of neighbors, the
density ratio between the natural and post-intervention exposure could grow
extremely large, destabilizing downstream estimates. More modern ``balancing''
tools, such as Riesz regression~\citep{Chernozhukov2021}, may help overcome this
issue.

In addition, practitioners must know how the interference arises. This does
often occur: in our data analysis, the form of interference arose naturally as
a part of the scientific question that considered an intervention on
\textit{all} vehicles entering a given ZCTA, not just those registered there.
If an investigator has reason to suspect interference, they may also suspect
the underlying process by which it occurs. However, there are cases where an
investigator may not know the form. Although discussed in recent
work~\citep{hoshino2023causal, ohnishi2022degree}, this remains an avenue for
future research.

Other potential areas of future research involve extending the induced MTP to
more complex settings. In time-varying data, for
example, one might consider using an \textit{induced longitudinal}
MTP~\citep{diazNonparametricCausalEffects2023}, which would involve
sequential regression-based algorithms to account for summary measures of
exposures subject to time-varying confounding in a network profile that
may itself evolve across time. Further work overcoming the technical
limitations of estimating the effects of induced MTPs will be critical to
facilitate answering more complex causal inference questions in settings where
continuous exposures are measured in datasets exhibiting network interference.

\section*{Acknowledgments}

SVB was supported in part by grants from the National Institute of
Environmental Health Sciences (award no.~T32 ES007142) and the National Science
Foundation (award no.~DGE 2140743). The authors thank Rachel Nethery for her
help in initiating the motivating data analysis.

\vspace{1em}
\noindent
\textit{Conflicts of interest}: The authors have no conflicts of interest to
disclose.

\def\UrlBreaks{\do\/\do-}
\bibliography{refs}
\end{document}

\maketitle

%%%%%%%%%%%%%%%%%%%%%%%%%%%%%%%%%%%%%%%%%%%%%%%%%%%%%%%%%%%%%%%%%%%%%%%%%%%%%%%

These supplementary materials provide proofs of the results presented in the
main text, elaborations on some mathematical details, and extra figures and
tables for the simulations and data analysis studying the effects of zero-emission vehicle
uptake on NO\textsubscript{2} air pollution in California.  

\section{Proof preliminaries}\label{section:preliminaries}

As a shorthand, let $\A$ and $\Li$ denote the support of $A_i$ and $L_i$,
respectively. Furthermore, let $\A_i^s$ and $\Li_i^s$ denote the support of
$A_i^s$ and $L_i^s$ (which may differ depending on the unit $i$ as each unit
may have different neighbors and network edges). Denote by $\Ab^d$ the vector
$[d(A_i, L_i; \delta)]_{i=1}^n$ so that
\begin{equation}
    A_i^{s \circ d} = s_{F_i}(\Ab^d, \Lb)
\end{equation}
\noindent and similarly, $L_i^s = s_{F_i}(\Lb)$. For further brevity, we simplify the exposure summary to either $s(\ab)$ or $s_{F_i}(\ab)$ when its dependence on data other than its exposure argument (and set of neighbors, for the latter) is irrelevant to a proof.

We wish to identify the causal
estimand $\Psi_n(\Pf) = \E\Big[\frac{1}{n}\sum_{i=1}^n Y(A_i^{s \circ d})\Big]$.
We begin with a preliminary lemma that will be helpful for identification.

\begin{lemma}\label{lemma:change-of-variables} Change-of-variables for
  multidimensional probability densities.

\noindent If $s: \mathbb{R}^n \mapsto \mathbb{R}^k$ with $k \leq n$ is a
  \jrssb{locally Lipschitz} function with Jacobian $J_{\ab}s$, $\Ab$ is a random vector with
  density function $p_\Ab$, and $A^s = s(\Ab)$, then the density of $A_s$
  satisfies
  \begin{equation}
    \int\displaylimits_{(\A)^n} p_{A^s}(s(\ab)) \sqrt{\det J_{\ab}s(\ab)J_{\ab}s(\ab)^\top}
    \partial \ab = \int\displaylimits_{ a^s \in \A^s}\int
    \displaylimits_{\substack{s(\ab) = a^s : \\ \ab \in (\A)^n}} p_\Ab(\ab)
    \partial \mu(\ab)\partial \mu(a^s) \ ,
  \end{equation}
\noindent where $\mu$ is the Hausdorff measure, which corresponds to the
Lebesgue measure on dense subsets of $\mathbb{R}^n$. In other words,
$p_{A^s}(s(\ab)) \sqrt{\det J_{\ab}s(\ab)J_{\ab}s(\ab)^\top}$ is the density of $p_A$ over
level sets of $s$.
\end{lemma}

\begin{proof}
  This lemma is a direct application of the \textit{coarea formula}, a
  generalization of the area formula from multivariable calculus to functions
  whose Jacobian matrices are not square. The measure-theoretic coarea formula
  was first proven by \cite{Federer1959, Federer1969}, but \cite{Negro2022}
  describes its contemporary use for transformations of continuous random
  variables. Applying Theorem 3.3 part (ii) with Definition 3.1 of
  \cite{Negro2022} to the setup given by the above, we have that
  \begin{align*}
   \int\displaylimits_{ a^s \in \A^s}\int\displaylimits_{\substack{s(\ab) =
   a^s : \\ \ab \in (\A)^n}} p_\Ab(\ab) \partial \mu(\ab)\partial \mu(a^s) &=
   \int\displaylimits_{ a^s \in \A^s}\int\displaylimits_{\substack{s(\ab) =
   a^s : \\ \ab \in (\A)^n}} p_{A^s}(s(\ab)) \partial \mu(\ab)\partial \mu(a^s)\\
   &= \int\displaylimits_{(\A)^n} p_{A^s}(s(\ab))
   \sqrt{\det J_{\ab}s(\ab)J_{\ab}s(\ab)^\top}\partial \ab \ .
  \end{align*}
  Specifically, the second term follows from the fact that integrating
  $p_\Ab$ over the level sets of $s$ is equivalent to integrating over
  $p_{A^s}$, and the third from Theorem 3.3 of \cite{Negro2022}. Interested
  readers may also consult Lemma 4.4 of \cite{Negro2022}, which provides the
  density of $p_{A^s}$ directly as a function of $p_\Ab(\ab)$ and $J_{\ab}s(\ab)$
  (that is, the reverse of this lemma).
\end{proof}

\jrssb{Given the assumptions established in the main text, an important remark will permit identification to proceed:}

\begin{remark}\label{remark:psi}
    \jrssb{The piecewise smooth invertibility assumption \ref{main:assumption:psi} and the coarea assumption~\ref{main:assumption:coarea} imply that, in an induced MTP, both $d^{-1}$ and $s$ are locally Lipschitz, and therefore $s \circ d^{-1}$ must be locally Lipschitz as well. This implies that Lemma~\ref{lemma:change-of-variables} can be applied to $s \circ d^{-1}$.}
\end{remark}

\noindent

\section{Identification}\label{section:identification}

In this section, we use the assumptions defined in the main manuscript to
identify the desired causal quantity $\Psi_n(\Pf)$ via a statistical functional
$\psi_n$ that is estimable under the observed data structure. First, we can apply iterated
expectation to note
\begin{align*}
    \psi_n &= \E_\Pf\Big(\frac{1}{n}\sum_{i=1}^nY(A_i^{s\circ d})\Big) 
    = \frac{1}{n}\sum_{i=1}^n \E_{Y(A_i^{s\circ d})}(\E_{\Lb}(Y(A_i^{s\circ d}) \mid \Lb))\tag{Iterated expectation}\\
    &= \frac{1}{n}\sum_{i=1}^n \E_{Y(A_i^{s\circ d})}(\E_{A_i^{s\circ d}, \Lb}(Y(A_i^{s\circ d}) \mid A_i^{s\circ d}, \Lb))\tag{No unmeasured confounding}\\
    &= \frac{1}{n}\sum_{i=1}^n \E_Y(\E_{A_i^{s\circ d}, L_i^s}(Y \mid A_i^{s\circ d}, L_i^s)) \tag{SCM: $Y$ only depends on $\Lb$ through $L_i^s$}
\end{align*}
\noindent
While technically a statistical quantity in that the final line does not depend
on counterfactual distributions of $Y$, in this problem we only observe $\Ab$
and $\Lb$, not $A_i^{s \circ d}$ and $L_i^s$. To ensure that this statistical
quantity is well-defined, we again apply an iterated expectation over $\Ab$ and
$\Lb$, both of which we do observe. Without loss of generality, assume $\Ab$
and $\Lb$ are continuous with density $p(\ab, \lb) = p(\ab \mid \lb)p(\lb)$. Also, for brevity, let $h_k(\ab) = [h_k(a_i)]_{i=1}^n$, and denote  the shorthand for the
determinant $\Delta f(\ab) = \sqrt{\det J_\ab f(\ab)J_\ab
f(\ab)^\top}$. Then:
\jrssb{\begin{align*}
    & \frac{1}{n}\sum_{i=1}^n \E_Y(\E_{A_i^{s\circ d}, L_i^s}(Y \mid A_i^{s\circ d}, L_i^s))\\
    &= \frac{1}{n}\sum_{i=1}^n \int\displaylimits_{ l^s \in \Li_i^s}\int\displaylimits_{\substack{s_{F_i}(\lb) = l^s : \\ \lb \in (\Li)^n}}\int\displaylimits_{ a^{s} \in \A_i^s}\int\displaylimits_{\substack{s_{F_i}(\ab) = a^{s} : \\ \ab \in (\A)^n}} m\Big((s_{F_i} \circ d)(\ab), s_{F_i}(\lb)\Big) p(\ab, \lb) \partial \mu(\ab)\partial\mu(a^s) \partial \mu(\lb )\partial\mu(l^s)\tag{Positivity} \\
    &= \frac{1}{n}\sum_{i=1}^n \int\displaylimits_{ l^s \in \Li_i^s}\int\displaylimits_{\substack{s(\lb) = l^s : \\ \lb \in (\Li)^n}}\sum_{k=1}^K\,\int\displaylimits_{ a^{s} \in \A_i^s}\int\displaylimits_{\substack{(s_{F_i}\circ h_k)(\ab) = a^{s} : \\ \ab \in (\A_k(l))^n}} m\Big(s_{F_i}(\ab), s_{F_i}(\lb)\Big) p(\ab, \lb) \partial \mu(\ab)\partial\mu(a^s) \partial \mu(\lb )\partial\mu(l^s)\tag{Change of level set using Assumption \ref{main:assumption:psi}} \\
    &= \frac{1}{n}\sum_{i=1}^n \int\displaylimits_{(\Li)^n}\sum_{k=1}^K\,\int\displaylimits_{(\A_k(l))^n} m(s_{F_i}(\ab), s_{F_i}(\lb)) \cdot p_{A^s}((s_{F_i} \circ h_k)(\ab) \mid s_{F_i}(\lb))\Delta (s_{F_i}\circ h_k)(\ab) \partial \ab \cdot p_{L^s}(s(\lb))\Delta s(\lb) \partial \lb \tag{Lemma \ref{lemma:change-of-variables} and Remark \ref{remark:psi}} \\
\end{align*}}
The first step follows from positivity because (i) the positivity assumption
guarantees $(s_{F_i} \circ d) \in \A^s$, and (ii) $m$ is only a function of
summaries, so we only need to integrate over values of the summary with
positive probability. \jrssb{In the second step, we note that integrating a function of $(s_{F_i} \circ d)(\ab)$ over the level set of $s(\ab) = a^s$ is equivalent to integrating a function of $s_{F_i}(\ab)$ over the level set $(s \circ d^{-1})(\ab) = a^s$, provided $d^{-1}$ exists; since $d$ is assumed piecewise smooth invertible, we can perform this change of level set on each subset $\A_k \in \A$ as defined in Assumption \ref{main:assumption:psi} using the piecewise inverse $h_k$. In the third step, we apply Lemma~\ref{lemma:change-of-variables} to the function $s \circ h_k$ following Remark~\ref{remark:psi}}

\jrssb{Next, we multiply and divide by the density of $A^s$ to convert the inner integral to a functional of the observed data and reconstruct the full integral:}
\jrssb{\begin{align*}
    & \sum_{k=1}^K\,\int\displaylimits_{(\A_k(l))^n} m(s_{F_i}(\ab), s_{F_i}(\lb)) \cdot p_{A^s}((s_{F_i} \circ h_k)(\ab) \mid s_{F_i}(\lb))\Delta (s_{F_i}\circ h_k)(\ab) \partial \ab \\
    &= \int\displaylimits_{(\A)^n} m(s_{F_i}(\ab), s_{F_i}(\lb)) \cdot \sum_{k=1}^K\, \I(\ab \in (\A_k(l))^n) p_{A^s}((s_{F_i} \circ h_k)(\ab) \mid s_{F_i}(\lb))\Delta (s_{F_i}\circ h_k)(\ab) \partial \ab\\
    &= \int\displaylimits_{(\A)^n} m(s_{F_i}(\ab), s_{F_i}(\lb)) \cdot \frac{p_{A^s}((s_{F_i} \circ d^{-1})(\ab) \mid s_{F_i}(\lb))\Delta (s_{F_i}\circ d^{-1})(\ab)}{p_{A^s}(s_{F_i}(\ab) \mid s_{F_i}(\lb))\Delta s_{F_i}(\ab)} p_{A^s}(s_{F_i}(\ab) \mid s_{F_i}(\lb))\Delta s_{F_i}(\ab)\partial \ab\\
    &= \int\displaylimits_{(\A)^n} m(a^s, l^s) \cdot r(a^s, a^{s \circ d^{-1}}, l^s) \cdot w(\ab, \lb, i) \cdot p(a^s \mid l^s)\Delta s_{F_i}(\ab, \lb)\partial \ab \ ,
\end{align*}}
\noindent
where the final integral notation replaces the summary functions over $\ab$ and $\lb$
with their shorthand $a^s$ and $l^s$ to emphasize that the nuisances $m$ and
$r$ only depend on the values of the summarized variables. \jrssb{Note the use of the notation $p_{A^s}((s_{F_i} \circ d^{-1})(\ab) \mid s_{F_i}(\lb))\Delta (s_{F_i}\circ d^{-1})(\ab)$, which by Assumption~\ref{main:assumption:psi} in the main text will equal the piecewise density
$\sum_{k=1}^K\, \I(\ab \in (\A_k)^n) p_{A^s}((s_{F_i} \circ h_k)(\ab) \mid s_{F_i}(\lb))\Delta (s_{F_i}\circ h_k)(\ab)$}. Assumptions~\ref{main:assumption:psi} and~\ref{main:assumption:coarea}
guarantee that $r$ and $w$ are well-defined almost everywhere. With the inner integral replaced,
the statistical estimand becomes
\begin{equation}
   \frac{1}{n}\sum_{i=1}^n \E_\Pf(m(A_i^s, L_i^s) \cdot r(A_i^s,
   A_i^{s\circ d}, L_i^s) \cdot w(\Ab, \Lb, i)) \ ,
\end{equation}
\noindent with nuisances denoted, taking $d^{-1}(\ab) = [d^{-1}(a_i)]_{i=1}^n$ and $d^{-1}(a_i)$ defined in Assumption \ref{main:assumption:psi},

\begin{align*}
    m(a^s, l^s) &= E_Y(Y \mid A^s = a^s, L^s = l^s) \\
    r(a^s, a^{s \circ d^{-1}}, l^s) &= \frac{p_{A^s}(a^{s \circ d^{-1}} \mid l^s)}{p_{A^s}(a^s \mid l^s)} \\
    w(\ab, \lb, i) &= \frac{\Delta (s_{F_i} \circ d^{-1})(\ab, \lb; \delta)}{\Delta s_{F_i}(\ab, \lb)} \ .
\end{align*}

Because only a single realization of the network $\Fb$ is observed, identifying
the estimand in terms of $m(A_i^s, L_i^s)$, $r(A_i^s, A_i^{s \circ d^{-1}},
L_i^s)$, and $w(\Ab, \Lb)$ is necessary for $\psi_n$ to be estimable from the
data. Note that only the first two nuisance quantities $m$ and $r$ depend on the
summaries: $w$ is a known function of only $s$ and $d$, which are specified by
the investigator, and the observed data. Consequently, only the first two
nuisances need to be estimated; one can do so non-parametrically by fitting
machine learning algorithms based on the summaries $A_i^s$ and $L_i^s$, rather
than needing to estimate the joint density of every individual exposure $\Ab$.

\section{Obtaining the efficient influence function}\label{section:eif}

Identification in terms of $w$ and $r$ also permits existing theory to be used
for estimation based on the efficient influence function. Recall that a
modified treatment policy is a type of stochastic intervention---an
intervention that replaces $P$ with some new distribution $P^\star$. The EIF of
a stochastic intervention in the network setting
of~\cite{ogburnCausalInferenceSocial2022} is
\begin{equation}
  \frac{1}{n}\sum_{i=1}^n\E(m(A_i^s, L_i^s) \mid \Lb) +
    \frac{\bar{p}^\star(A_i^s, L_i^s)}{\bar{p}(A_i^s, L_i^s)}
    (Y - m(A_i^s, L_i^s)) - \psi_n \ ,
\end{equation}
where the weighting function for the stochastic intervention is
\begin{equation}
\frac{\bar{p}^\star(A_i^s, L_i^s)}{\bar{p}(A_i^s, L_i^s)} =
  \frac{\frac{1}{n}\sum_{j=1}^nP^\star(A_j^s = a_i^s, L_j^s = l_i^s)}
  {\frac{1}{n}\sum_{j=1}^nP(A_j^s = a_i^s, L_j^s = l_i^s)}
\end{equation}

\noindent with $P^\star$ denoting the probability measure of the replacement
distribution. The numerator and denominator are denoted as means because they
represent mixture distributions of summaries but are, in practice, estimated
together using a single density estimator
\citep{zivichTargetedMaximumLikelihood2022}. In our setting, these can be
simplified by the law of total probability to marginalize over vectors of
neighbors:

\begin{align*}
\frac{1}{n}\sum_{j=1}^nP(A_j^s = a_i^s, L_j^s = l_i^s) &= \frac{1}{n}\sum_{j=1}^nP(s_{F_j}(\Ab, \Lb) = a_i^s, s_{F_j}(\Lb) = l_i^s \mid F_j) \\
&= \sum_{j=1}^nP(s_{F_j}(\Ab, \Lb) = a_i^s, s_{F_j}(\Lb) = l_i^s \mid F_j)\frac{\mathbb{I}(F_j = \mathbf{f}_j)}{n}\\
&= \sum_{j=1}^nP(s_{F_j}(\Ab, \Lb) = a_i^s, s_{F_j}(\Lb) = l_i^s \mid F_j)P(F_j)\\
&= P(A^s = a_i^s, L^s = l_i^s) \tag{Law of total probability}\\
&= P(A^s = a_i^s \mid L^s = l_i^s) P(L^s = l_i^s) \tag{Law of conditional probability}
\end{align*}

\noindent
where the final probability is the marginal density of $A^s$ and $L^s$ over all
possible $F_j$. The same equality holds for $P^\star$. Hence, we can simply
apply the same change-of-variables from our identification result to obtain
\begin{align*}
  \frac{P^\star(A^s = a_i^s, L^s = l_i^s)}{P(A^s = a_i^s, L^s = l_i^s)} =
  \frac{P^\star(A^s = a_i^s \mid L^s = l_i^s)\cancel{P(L^s = l_i^s)}}
       {P(A^s = a_i^s \mid L^s = l_i^s)\cancel{P(L^s = l_i^s )}} \\
  = \frac{p_{A^s}(A_i^{s \circ d^{-1}} \mid L_i^s)}
         {p_{A^s}(A_i^s \mid L_i^s)}\frac{\Delta (s_{F_i} \circ
         d^{-1})(\Ab, \Lb; \delta))}{\Delta s_{F_i}(\Ab, \Lb)}
  = r(A_i^s, A_i^{s \circ d^{-1}}, L_i^s)w(\Ab, \Lb) \ .
\end{align*}

\noindent
Plugging this into the EIF from~\cite{ogburnCausalInferenceSocial2022}---also
derived in \cite{sofryginSemiParametricEstimationInference2017}---we obtain
the EIF for the population intervention effect of an MTP on network $\Fb$:
\begin{equation}
  \frac{1}{n}\sum_{i=1}^n \Big(r(A_i^s, A_i^{s \circ d^{-1}}, L_i^s)
    w(\Ab, \Lb)(Y_i - m(A^s_i, L_i^s)) +
  \E_\Pf(m(A^{s\circ d}_i, L_i^s) \mid \Lb = \lb)\Big) - \psi_n \ .
\end{equation}

\section{One-step estimator}\label{section:one-step}

Here, we prove that the one-step estimator given by
Equation~\eqref{main:eq:one-step} in the main manuscript solves the EIF estimating
equation asymptotically and is therefore consistent and semi-parametric
efficient under the same estimator regularity conditions as given
by~\cite{ogburnCausalInferenceSocial2022}. As a reminder, for data $O_i$ with a
network $F$ admitting a CLT of rate $C_n$, these regularity conditions are:

\begin{assumption}\label{assumption:product-rate}
  \textit{Product rate convergence}: $\lVert \hat{m} - m \rVert_2 \lVert
    \hat{r} - r \rVert_2 = o_{\Pf}(C_n^{-1/2})$ \ .
\end{assumption}

\begin{assumption}\label{assumption:empirical-process}
  \textit{Empirical process condition}: $\frac{1}{n}\sum_{i=1}^n
  \phi_{\hat{\Pf}_n}(O_i) - \phi_{\Pf_n}(O_i) -
  \E_\Pf(\phi_{\hat{\Pf}_n}(O_i) - \phi_{\Pf_n}(O_i)) = o_\Pf(C_n^{-1/2})$ \ .
\end{assumption}

Assumption~\ref{assumption:product-rate} can be satisfied by
suitable choices of nuisance estimators. We ensure
Assumption~\ref{assumption:empirical-process} using
cross-fitting; see Section~\ref{section:crossfitting}
for further details. Under these assumptions, the following theorem holds:

\begin{theorem}
    Under assumptions \ref{assumption:product-rate} and \ref{assumption:empirical-process}, the simplified one-step estimator is consistent: 
    \begin{equation}
        \frac{1}{n}\sum_{i=1}^n \phi_{\hat{\Pf}_n}(O_i)  \overset{p} {\rightarrow} \psi_n
    \end{equation}
\end{theorem}

\begin{proof}
Solving the EIF in Equation~\eqref{main:eq:eif-equation} directly for $\psi_n$
yields the quantity
\begin{equation}
  \psi_n = \frac{1}{n}\sum_{i=1}^n \hat{r}(A_i^s, A_i^{s \circ d^{-1}}, L_i^s)
    w(\Ab, \Lb)(Y_i -
  \hat{m}(A_i^s, L_i^s)) + \E_{\Pf}(\hat{m}(h(A_i^s,L_i^s; \delta),L_i^s)
  \mid \Lb = \lb) \ .
\end{equation}
Distributing the summation linearly, this can be divided into two components,
which both converge asymptotically by Theorem~\ref{main:theorem:ogburn}. The first is the reweighted residual, 
\begin{align*}
  \frac{1}{n}\sum_{i=1}^n &\hat{r}(A_i^s, A_i^{s \circ d^{-1}}, L_i^s)
    w(\Ab, \Lb)(Y_i - \hat{m}(A_i^s, L_i^s))\\
  & \overset{p}{\rightarrow} \E_\Pf(r(A_i^s, A_i^{s \circ d^{-1}}, L_i^s)
    w(\Ab, \Lb)(Y_i - \hat{m}(A_i^s, L_i^s))) \ ,
\end{align*}

\noindent which corrects the bias of
the plug-in estimate. The second is the plug-in estimate itself,

\begin{align*}
  \frac{1}{n}\sum_{i=1}^n\E_{\mathbf{A}|\mathbf{L}}(\hat{m}(h(A_i^s, L_i^s; \delta), L_i^s)
  \mid \Lb = \lb) &\overset{p}{\rightarrow}\E_\Lb(\E_{\mathbf{A}|\mathbf{L}}(\hat{m}(h(A_i^s, L_i^s;
  \delta), L_i^s) \mid \Lb = \lb)) \\
  &=  \E_{\Lb, \Ab}(\hat{m}(h(A_i^s, L_i^s; \delta)) \ ,
\end{align*}
where the third line follows from the law of total expectation. But, recall that, equivalently, the direct plug-in estimator also converges (assuming $\hat{m}$ is consistent):
\begin{equation}
  \frac{1}{n}\sum_{i=1}^n\hat{m}(h(A_i^s, L_i^s; \delta))\overset{p}
  {\rightarrow}\E_{\Lb, \Ab}(\hat{m}(h(A_i^s, L_i^s; \delta), L_i^s)) \ .
\end{equation}
Therefore, by the continuous mapping theorem,
\begin{align*}
  \hat{\psi}_n^{\text{OS}} = \frac{1}{n}\sum_{i=1}^n\hat{r}(A_i^s, A_i^{s \circ d^{-1}}, L_i^s)
    w(\Ab, \Lb)
  (Y_i - \hat{m}(A_i^s,L_i^s)) + \frac{1}{n}\sum_{i=1}^n\hat{m}(h(A_i^s, L_i^s;
  \delta), L_i^s) \\\overset{p}{\rightarrow} \E_\Pf(\hat{r}(A_i^s, A_i^{s \circ d^{-1}}, L_i^s)
    w(\Ab, \Lb)(Y_i -
  \hat{m}(A_i^s, L_i^s))) + \E_{\Lb, \Ab}(\hat{m}(h(A_i^s, L_i^s; \delta))
  = \psi_n \ ,
\end{align*}
proving that the one-step solves the same estimating equation as
Equation~\eqref{main:eq:eif} asymptotically.
\end{proof}

\section{Derivation of the variance estimator}\label{section:variance-estimator}

In this section we prove that $\hat{\sigma}^2$ as defined in
Section~\ref{main:section:variance} is a consistent estimator of the variance
of $\hat{\psi}_n^{\text{OS}}$, and by extension, the variance of
$\hat{\psi}_n^{\text{TMLE}}$. To do this, we establish a few supporting lemmas that hold under the assumptions of the main paper. The first supporting lemma concerns the consistency of the one-step estimator within individual strata of possible $|F_i|$ values; this will constitute an important component of the variance estimator later.

\begin{lemma}[Consistency of $\hat{\psi}(|F_i|)$]\label{lemma:sub-consistent}
A one-step estimator applied to the subset of units with number of friends $|F_i|$ is quarter-root consistent: that is,
$\hat{\psi}(|F_i|) - \E(\phi_{\hat{\Pf}_n}(O_i)) =
o_\Pf(n^{-1/4})$.
\end{lemma}

\begin{proof}
In Section~\ref{main:section:identification} we assumed that
$P(A_i^s) > 0$ for all $i$. For this assumption to hold true, the number of
neighbors of any unit $i$ cannot grow asymptotically slower than that of any
other unit $j$. Formally, this means $|\mathcal{N}(|F_i|)| \propto
|\mathcal{N}(|F_j|)|$ for all $i,j \in 1,\ldots, n$.

Next, recall that the convergence of the one-step estimator assumes that the
maximum degree of any node in the network encoded by $\mathbf{F}$ grows no
faster than $n^{1/2}$. More formally, this means $\max_i |F_i| = O(n^{1/2})$.
Since $|\mathcal{N}(|F_i|)| \propto |\mathcal{N}(|F_j|)|$, for all $i,j$,
noting that $n / n^{1/2} = n^{1/2}$, it must be true that $|\mathcal{N}(|F_i|)|
= O(n^{1/2})$ for all $i$. Then, noting that under
Assumptions~\ref{assumption:product-rate}
and~\ref{assumption:empirical-process},
$\hat{\psi}_n^{\text{OS}} - \psi_n = o_\Pf(n^{-1/2})$ for networks of bounded
degree, computing the one-step estimator on the subset of nodes
$\mathcal{N}(|F_i|)$, which have identical and therefore bounded degree, yields
\begin{equation}
  \hat{\psi}_n(|F_i|) - \E(\phi_{\hat{\Pf}_n}(O_i)) =
  o_\Pf(|\mathcal{N}(|F_i|)|^{-1/2}) =
  o_\Pf(O(n^{1/2})^{-1/2}) = o_\Pf(n^{-1/4}) \ .
\end{equation}
This completes the proof that the one-step estimator $\hat{\psi}(|F_i|)$
converges to $\E(\phi_{\hat{\Pf}_n}(O_i))$, specifically at rate
$o_\Pf(n^{-1/4})$.
\end{proof}

Next, we prove a lemma demonstrating that the true variance of the centered estimating function is, in fact, the variance that we want to estimate. 

\begin{lemma}[Bias and variance of estimating function]\label{lemma:est-func}
The estimating function $\phi_{\hat{\Pf}_n}(O_i) -
\hat{\psi}_n(|F_i|)$ is unbiased -- that is, $E(\phi_{\hat{\Pf}_n}(O_i) -
\hat{\psi}_n(|F_i|)) = 0$ -- and, subsequently, its variance is equal to that of the
one-step estimator: $\text{Var}(\phi_{\hat{\Pf}_n}(O_i) -
\hat{\psi}_n(|F_i|)) = \text{Var}(\hat{\psi}_n^{\text{OS}}) = \sigma^2$
\end{lemma}

\begin{proof}
First, note that under the assumed SCM~\eqref{main:npsem2}
in Section~\ref{main:section:identification}, $\phi_{\hat{\Pf}_n}(O_i)$ and
$\phi_{\hat{\Pf}_n}(O_j)$ are identically distributed if $|F_i| = |F_j|$.
Consequently,
\begin{equation}
  \E(\phi_{\hat{\Pf}_n}(O_i) - \hat{\psi}_n(|F_i|)) =
  \E(\phi_{\hat{\Pf}_n}(O_i)) - \frac{1}{\mathcal{N}(|F_i|)}
  \sum_{j \in \mathcal{N}(|F_i|)}\E(\phi_{\hat{\Pf}_n}(O_j)) =
  \E(\phi_{\hat{\Pf}_n}(O_i)) - \E(\phi_{\hat{\Pf}_n}(O_j)) \ .
\end{equation}
Since $|F_i| = |F_j|$ by definition of $\mathcal{N}(|F_i|)$,
$\E(\phi_{\hat{\Pf}_n}(O_i)) - \E(\phi_{\hat{\Pf}_n}(O_j)) = 0$. Secondly, by
mathematical properties of the variance, the variance of our estimating equation
can be rewritten as follows:
\begin{align*}
  \text{Var}\Big(\frac{1}{n}\sum_{i=1}^n \phi_{\hat{\Pf}_n}(O_i) - \psi\Big) &=
  \frac{1}{n^2}\sum_{i,j} \E\Big((\phi_{\hat{\Pf}_n}(O_i) - \E(\phi_{\hat{\Pf}_n}(O_i)))
  (\phi_{\hat{\Pf}_n}(O_j) - \E(\phi_{\hat{\Pf}_n}(O_j)))\Big)\\
  &=\text{Var}\Big(\frac{1}{n}\sum_{i=1}^n(\phi_{\hat{\Pf}_n}(O_i) -
    \E(\phi_{\hat{\Pf}_n}(O_i)))\Big) = \sigma^2 \ ,
\end{align*}

\noindent the variance of $\hat{\psi}_n^{\text{OS}}$ that we want to estimate, which
completes the proof. 
\end{proof}

\begin{theorem}[Plug-in variance consistency]
The plug-in variance estimator is consistent:
\begin{equation}
  \hat{\sigma}^2 = \frac{1}{n^2}\sum_{i,j}G(i, j) \varphi_i \varphi_j \overset{p}{\rightarrow} \sigma^2\ ,
\end{equation}
where $\varphi_i = \phi_{\hat{\Pf}_n}(O_i) - \hat{\psi}_n(|F_i|)$. 
\end{theorem}
\begin{proof}
By adding
and subtracting to obtain the equality
\begin{equation*}
    \phi_{\hat{\Pf}_n}(O_i) - \hat{\psi}_n(|F_i|) = \phi_{\hat{\Pf}_n}(O_i) -
    \E(\phi_{\hat{\Pf}_n}(O_i)) + \E(\phi_{\hat{\Pf}_n}(O_i)) -
    \hat{\psi}_n(|F_i|) \ ,
\end{equation*}
we can decompose this estimator as follows:
\begin{align}\label{eq:var-decomposition}
  \hat{\sigma}^2 =& \frac{1}{n^2}\sum_{i,j}G(i, j)\Big(\phi_{\hat{\Pf}_n}(O_i) - \hat{\psi}_n
    (|F_i|)\Big)\Big(\phi_{\hat{\Pf}_n}(O_j) - \hat{\psi}_n(|F_j|)\Big)\nonumber\\
  =& \frac{1}{n^2}\sum_{i,j}G(i, j)\Big(\phi_{\hat{\Pf}_n}(O_i) - \E(\phi_{\hat{\Pf}_n}(O_i))\Big)\Big(\phi_{\hat{\Pf}_n}(O_j) - \E(\phi_{\hat{\Pf}_n}(O_j))\Big)\\
  &- \frac{2}{n^2}\sum_{i,j}G(i, j)\Big(\phi_{\hat{\Pf}_n}(O_i) - \E(\phi_{\hat{\Pf}_n}(O_i))\Big)\Big(\hat{\psi}_n(|F_j|) -
    \E(\phi_{\hat{\Pf}_n}(O_j))\Big) \nonumber\\
  &+ \frac{1}{n^2}\sum_{i,j}G(i, j)\Big(\hat{\psi}_n(|F_i|) -
    \E(\phi_{\hat{\Pf}_n}(O_i))\Big)\Big(\hat{\psi}_n(|F_j|) - \E(\phi_{\hat{\Pf}_n}(O_j))\Big)\nonumber \ .
\end{align}
From~\cite{ogburnCausalInferenceSocial2022}, we know that
\begin{equation}
\frac{1}{n^2}\sum_{i,j}G(i, j)\Big(\phi_{\hat{\Pf}_n}(O_i) -
  \E(\phi_{\hat{\Pf}_n}(O_i))\Big)\Big(\phi_{\hat{\Pf}_n}(O_j) -
  \E(\phi_{\hat{\Pf}_n}(O_j))\Big)\overset{p}{\rightarrow} \sigma^2 \ .
\end{equation}
which by Lemma \ref{lemma:est-func} is the desired variance to be estimated. Furthermore, we know that the one-step estimator is $\sqrt{C_n}$-consistent,
and since in Lemma~\ref{lemma:sub-consistent}
we proved that $\hat{\psi}_n(|F_j|) - \E(\phi_{\hat{\Pf}_n}(O_j)) =
o_\Pf(n^{-1/4})$, by the continuous mapping theorem,
\begin{align}
    & \frac{1}{n^2}\sum_{i,j}G(i, j)\Big(\phi_{\hat{\Pf}_n}(O_i) -
      \E(\phi_{\hat{\Pf}_n}(O_i))\Big)\Big(\hat{\psi}_n(|F_j|) - \E(\phi_{\hat{\Pf}_n}(O_j))\Big) \nonumber\\
    &= \Big(\frac{1}{n}\sum_{i}(\phi_{\hat{\Pf}_n}(O_i) - \E(\phi_{\hat{\Pf}_n}(O_i)))\Big)
      \Big(\frac{1}{n}\sum_{j}(\hat{\psi}_n(|F_j|) - \E(\phi_{\hat{\Pf}_n}(O_j)))\Big) \\
    &= o_\Pf(C_n^{-1/2}) \cdot o_\Pf(n^{-1/4}) = o_\Pf(C_n^{-1/2} \cdot
      n^{-1/4})\nonumber
\end{align}
which is faster than $C_n^{-1/2}$, and also
\begin{align}
  & \frac{1}{n^2}\sum_{i,j}G(i, j)(\hat{\psi}_n(|F_i|) -
    \E(\phi_{\hat{\Pf}_n}(O_i)))(\hat{\psi}_n (|F_j|) - \E(\phi_{\hat{\Pf}_n}(O_j))) \nonumber\\
    &= \Big(\frac{1}{n}\sum_{i}(\hat{\psi}_n(|F_i|) -
    \E(\phi_{\hat{\Pf}_n}(O_i)))\Big)\Big(\frac{1}{n}\sum_{j}(\hat{\psi}_n(|F_j|) -
    \E(\phi_{\hat{\Pf}_n}(O_j)))\Big) \\
  &= o_\Pf(n^{-1/4}) \cdot o_\Pf(n^{-1/4}) = o_\Pf(n^{-1/2})\nonumber \ ,
\end{align}
implying that the second, third, and fourth terms in
Equation~\eqref{eq:var-decomposition} converge to zero. Since
the first term converges in probability to $\sigma^2$, we have thereby proven
that $\hat{\sigma}^2 \overset{p}{\rightarrow} \sigma^2$.
\end{proof}

\section{Sample-splitting and cross-fitting}\label{section:crossfitting}

Here we establish that the typical as-IID sample splitting and cross-fitting
procedure remains valid even in our established network setting. Recall our
estimating equation is
\begin{equation}
  \frac{1}{n}\sum_{i=1}^n \phi_{\hat{\Pf}_n}(O_i) - \psi_n \ ,
\end{equation}
which decomposes into the form
\begin{align*}
  \Big(\frac{1}{n}\sum_{i=1}^n \phi_{\Pf_n}(O_i) - \E_\Pf(\phi_{\Pf_n}(O_i))\Big) +
  \Big(\E_\Pf(\phi_{\hat{\Pf}_n}(O_i) - \phi_{\Pf_n}(O_i))\Big) \\
  + \Big(\frac{1}{n}\sum_{i=1}^n (\phi_{\hat{\Pf}_n}(O_i) - \phi_{\Pf_n}(O_i)) -
    \E_\Pf(\phi_{\hat{\Pf}_n}(O_i) - \phi_{\Pf_n}(O_i))\Big) \ .
\end{align*}
In standard semi-parametric theory, the first term is controlled by the CLT,
the second by whether the nuisance estimators converge, and the third by
sample-splitting or by invoking empirical process theory. So far, we have used
theory from~\cite{ogburnCausalInferenceSocial2022}---what remains is only to
determine whether the third ``empirical process'' term can be controlled by
cross-fitting.

\noindent
Essentially, what this requires is that, if we knew the true estimating
function $\phi_{\Pf_n}$, the sample mean constructed as $\frac{1}{n}\sum_{i}
(\phi_{\hat{\Pf}_n}(O_i) - \phi_{\Pf_n}(O_i))$ would be a \textbf{consistent}
estimator of the true bias of the estimating function at the correct rate. We
establish that this holds in our setting in the following theorem:

\begin{theorem}[As-IID Sample Splitting in Network Data]
    Let the assumptions of the CLT of~\cite{ogburnCausalInferenceSocial2022}
    hold (Theorem~\ref{main:theorem:ogburn} in the main manuscript). Suppose we
    draw two samples of size $n$, or ``folds,'' of indices $S_1$ and $S_2$
    chosen randomly and independently (i.e., not dependent on the data), and
    with the nuisance estimators of $\hat{\Pf}_n$ estimated only using $O_i : i
    \in S_1$. Then,
    \begin{equation}
      \frac{1}{n}\sum_{i \in S_2} (\phi_{\hat{\Pf}_n}(O_i) - \phi_{\Pf_n}(O_i))
      - \E_\Pf(\phi_{\hat{\Pf}_n}(O_i) - \phi_{\Pf_n}(O_i)) = o_{\Pf}(C_n^{-1/2}) \ .
    \end{equation}
    \noindent This means the ``empirical process'' term is asymptotically
    negligible.
\end{theorem}

\begin{proof}
Recall that for a statistic to be $o_\Pf(C_n^{-1/2})$ requires two conditions: (1)
asymptotic unbiasedness and (2) variance converging to zero at rate $1/C_n$.

\textit{We start with (1) asymptotic unbiasedness}. Let $S$ denote the set of data over which the set of influence functions is summed. Then,
\begin{align*}
  \E_\Pf\Big(\frac{1}{n}\sum_{i \in S} (\phi_{\hat{\Pf}_n}(O_i) -
    \phi_{\Pf_n}(O_i))\Big) &= \frac{1}{n}\sum_{i \in S}
    \E_\Pf(\phi_{\hat{\Pf}_n}(O_i) -\phi_{\Pf_n}(O_i) \mid S)\\
    &= \sum_{k=0}^{K_{\max}} \E_\Pf(\phi_{\hat{\Pf}_n}(O_i) -\phi_{\Pf_n}(O_i) \mid  S, |F_i| = k)P(|F_i| = k)\\
    &= \E_\Pf(\phi_{\hat{\Pf}_n}(O_i) -\phi_{\Pf_n}(O_i) \mid  S)
\end{align*}

\noindent where the last line follows from the law of conditional expectation, as the summation integrates over all values of $K_{\max}$, even as $K_{\max} \rightarrow \infty$. Sample splitting becomes necessary when it may be possible that 
\begin{equation*}
    \E_\Pf(\phi_{\hat{\Pf}_n}(O_i) -\phi_{\Pf_n}(O_i) \mid  S) \neq \E_\Pf(\phi_{\hat{\Pf}_n}(O_i) -\phi_{\Pf_n}(O_i))
\end{equation*}
As an example, if $S = S_1$ and our nuisance estimators perfectly interpolated each point in
$S_1$, then the bias would be $\E_\Pf(\phi_{\hat{\Pf}_n}(O_i) -
\phi_{\Pf_n}(O_i) \mid S_1) = 0$ even though
$\E_\Pf(\phi_{\hat{\Pf}_n}(O_i) - \phi_{\Pf_n}(O_i)) \neq 0$ (overfitting).
But, if we instead compute the sum over $S_2$,  chosen randomly such that $S_2 \cap S_1 = \emptyset$, even if
the individual data in $S_2$ are correlated with the data in $S_1$, then by the same logic as the above we have
\begin{align*}
  \E_\Pf\Big(\frac{1}{n}\sum_{i \in S_2} (\phi_{\hat{\Pf}_n}(O_i) -
  \phi_{\Pf_n}(O_i))\Big) &= \E_\Pf(\phi_{\hat{\Pf}_n}(O_i) -
  \phi_{\Pf_n}(O_i) \mid S_2) \\&= 
  \E_\Pf(\phi_{\hat{\Pf}_n}(O_i) -\phi_{\Pf_n}(O_i)) \ ,
\end{align*}
\noindent
since the choice of indices $S_2$ is made independently of $S_1$ and the observed data,
and the sample mean is an unbiased estimator even in correlated data.

\textit{(2) Variance converging to zero}. Consequently, whether as-iid
sampling-splitting is still valid in the correlated data setting primarily
depends on whether its variance converges at the appropriate rate. Under sample
splitting, since we just proved that our bias estimator is unbiased,
\begin{align*}
    \text{Var}\Big(\frac{1}{n}\sum_{i \in S_2} \phi_{\hat{\Pf}_n}(O_i) -\phi_{\Pf_n}(O_i)\Big) = E\Big(\text{Var}\Big(\frac{1}{n}\sum_{i \in S_2} (\phi_{\hat{\Pf}_n}(O_i) -\phi_{\Pf_n}(O_i) \mid S_2\Big)\Big) \tag{law of total variance}\\
    = \E\Big(\frac{1}{n^2}\sum_{i,j \in S_2} \text{Cov}\Big(\phi_{\hat{\Pf}_n}(O_i) -\phi_{\Pf_n}(O_i), \phi_{\hat{\Pf}_n}(O_j) -\phi_{\Pf_n}(O_j) \mid S_2\Big)\Big) \tag{Bienaym\'{e}'s identity}  \ .
\end{align*}
\noindent By the triangle inequality, 
\begin{align*}
  \Big| \frac{1}{n^2}\sum_{i,j \in S_2}&\text{Cov}\Big(\phi_{\hat{\Pf}_n}(O_i) -\phi_{\Pf_n}(O_i), \phi_{\hat{\Pf}_n}(O_j) -\phi_{\Pf_n}(O_j) \mid S_2\Big)\Big| \\ \leq \frac{1}{n^2}\sum_{i,j \in S_2}& \Big|\text{Cov}\Big(\phi_{\hat{\Pf}_n}(O_i) -\phi_{\Pf_n}(O_i), \phi_{\hat{\Pf}_n}(O_j) -\phi_{\Pf_n}(O_j) \mid S_2\Big)\Big| \\
  = \frac{1}{n^2}\sum_{i,j \in S_2}& \Big|\text{Corr}(\phi_{\hat{\Pf}_n}(O_i) -\phi_{\Pf_n}(O_i), \phi_{\hat{\Pf}_n}(O_j) -\phi_{\Pf_n}(O_j) \mid S_2) \\&\cdot \text{Var}(\phi_{\hat{\Pf}_n}(O_i) -\phi_{\Pf_n}(O_i) \mid S_2)\cdot\text{Var}(\phi_{\hat{\Pf}_n}(O_j) -\phi_{\Pf_n}(O_j) \mid S_2)\Big| \ ,
\end{align*}
\noindent
so, since the variance of the estimating function under our nuisance
estimators must be bounded, we can apply the dominated convergence theorem to
conclude that the variance of our bias estimator converges to zero, \textit{provided that}
not too many units are correlated.

Having data with a limited number of correlated units is needed to ensure that the above
satisfies $\frac{1}{n^2}\sum_{i,j \in S_2}
\Big|\text{Cov}\Big(\phi_{\hat{\Pf}_n}(O_i) -\phi_{\Pf_n}(O_i),
\phi_{\hat{\Pf}_n}(O_j) -\phi_{\Pf_n}(O_j) \mid S_2\Big)\Big| = o_\Pf(1/C_n)$.
Fortunately, since we have assumed $K_{\max} = o(\sqrt{n})$, we have that
\begin{align*}
    & \frac{1}{n^2}\sum_{i,j \in S_2} \Big|\text{Cov}\Big(\phi_{\hat{\Pf}_n}(O_i) -\phi_{\Pf_n}(O_i), \phi_{\hat{\Pf}_n}(O_j) -\phi_{\Pf_n}(O_j) \mid S_2\Big)\Big|\\
    & \leq \frac{nK_{\max}^2}{n^2}\max_{i, j \in S_2}\Big|\text{Cov}\Big(\phi_{\hat{\Pf}_n}(O_i) -\phi_{\Pf_n}(O_i), \phi_{\hat{\Pf}_n}(O_j) -\phi_{\Pf_n}(O_j) \mid S_2\Big)\Big|\\
    &= o(K_{\max}^2 /n) = o(1/C_n) \ ,
\end{align*}
\noindent
because if every unit has $K_{\max}$ neighbors, they will be correlated with
$K^2_{\max}$ units (their neighbors and their neighbors' neighbors), and $n /
K_{\max}^2 \leq C_n$ by assumption in the CLT
of~\cite{ogburnCausalInferenceSocial2022}. From this, we conclude that the
variance of our empirical process term converges at $o_{\Pf}(1/C_n)$, the
necessary rate.

Finally, since $\frac{1}{n}\sum_{i \in S_2} (\phi_{\hat{\Pf}_n}(O_i) -
\phi_{\Pf_n}(O_i))$ is an unbiased estimator of $\E_\Pf(\phi_{\hat{\Pf}_n}(O_i)
- \phi_{\Pf_n}(O_i))$ whose variance converges to zero at rate $1/C_n$, we have
that the empirical process term is $o_\Pf(C_n^{-1/2})$; hence, as-IID sample
splitting still serves its purpose to eliminate the empirical process term in our network setting. The validity of
cross-fitting in the correlated data setting for ensuring a fast convergence
rate on the empirical process term follows directly from the validity of
sample-splitting.
\end{proof}

\section{Additional Experimental and Data Analysis Results}\label{section:data-sources}

In this section, we include additional details to supplement the simulation experiments and data analysis conducted in the main manuscript. 

\subsection{Further elaboration on the experimental efficiency bound and scaling factor $C_n$}\label{section:eff-bound-elaboration}

In the synthetic data experiments, specifically Figure \ref{main:fig:synthetic}, we evaluate performance in terms of the network-TMLE estimator's scaled MSE against a finite-sample efficiency bound. This bound represents the variance of the \textit{ground truth} EIF, computed using the true outcome regression and density ratio functions. For each network structure, we obtain this variance approximately for each possible sample size by drawing a dataset, computing the variance of the ground truth EIF, and then replicating this procedure 10,000 times and averaging the results. 

We do this for two reasons. First, as mentioned in the main text, our estimator implicitly conditions on the network structure and estimates a parameter that depends on the sample size, so any efficiency bound should as well. Second, because $K_{\max}$ for each network structure grows roughly at rate $\log(n)$, $C_n^{-1/2} \approx \log(n)/\sqrt{n}$, so the asymptotic efficiency scales with $\log(n)$; this $\log$ term ensures even a single very large sample (i.e. $n = 10^6$) cannot possibly reflect an accurate bound to be approached. In light of these, the variance of the asymptotically optimal estimator at each sample size will serve as a reasonable finite-sample efficiency bound, showing us how close to optimal our estimator becomes as $n$ increases. 

\subsection{Additional simulation results}\label{section:more-sim-results}

\begin{figure}[htp]
\centering
\includegraphics[scale=1]{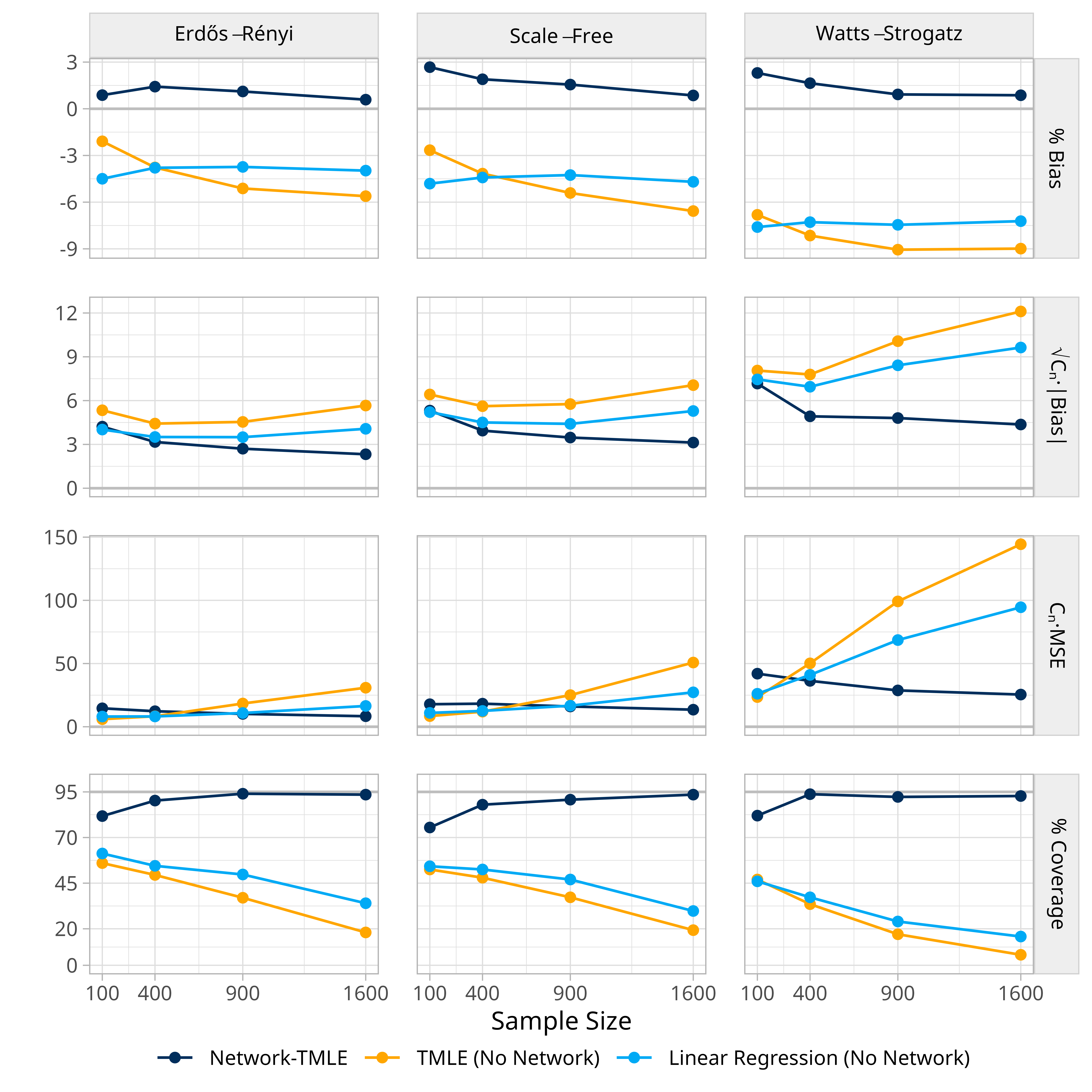}
\caption{Performance of network-TMLE compared to other common procedures for
  MTP estimation with various types of network profiles.}
\label{fig:synthetic-comparison}
\end{figure}

In this subsection, we include a few auxiliary simulation results to provide additional insights. First, Figure \ref{fig:synthetic-comparison} displays expanded results from the same synthetic data simulation used to create Figure \ref{main:fig:synthetic} in Section \ref{main:section:synthetic-data}, this time comparing network-TMLE to a classical ``IID'' TMLE and a linear regression, neither of which accounts for interference. Breaking out the results separately by network type, we can see that the classical methods incur more bias---and the scaled bias and MSE only increase with sample size. As a consequence, these methods suffer severe under-coverage, which also worsens as the number of samples grows. The gap in performance between network-TMLE and other methods appears largest for the Watts-Strogatz network, which features the greatest degree of connection between units among all of the simulated networks, and therefore the highest amount of interference.

Tables \ref{table:semisynthetic-big-mean} and \ref{table:semisynthetic-small-mean} depict results from additional semi-synthetic data simulations that differed from those of the main text only in the size of the individual exposure effects. In the Table \ref{table:semisynthetic-big-mean} simulation, the mean of the exposure was $\mu_A \approx 8.4$, so most units' exposure resulted in a positive effect. Conversely, in the Table \ref{table:semisynthetic-small-mean} simulation, the exposure mean was $\mu_A \approx 0.4$---an almost even mix of positive and negative exposures among individual units. 

For reference, the main text simulations, which had $\mu_A \approx 3.4$, displayed a simulation set-up where network-TMLE achieved uniformly better performance over both classical TMLE and linear regression, which both ignored interference. These two tables show how varying the magnitude of individual exposure effects can impact differences in performance. On one hand, when most exposures were positive (Table \ref{table:semisynthetic-big-mean}), classical methods incurred a much more severe bias and extremely low coverage. Network-TMLE corrects that bias, at the expense of a slight increase in variance compared to classical methods. On the other hand, under a mix of positive and negative exposures, classical methods incurred considerably less bias and much better coverage, but at the expense of a much larger variance than the network-TMLE. This ``bias-variance tradeoff'' shows how classical methods can fail in different ways when interference is ignored, depending on the treatment effect magnitude. 

\subsection{Data analysis summary statistics}
\begin{figure}[H]
\centering
\begin{subfigure}{.5\textwidth}
  \centering
  \includegraphics[scale=0.6]{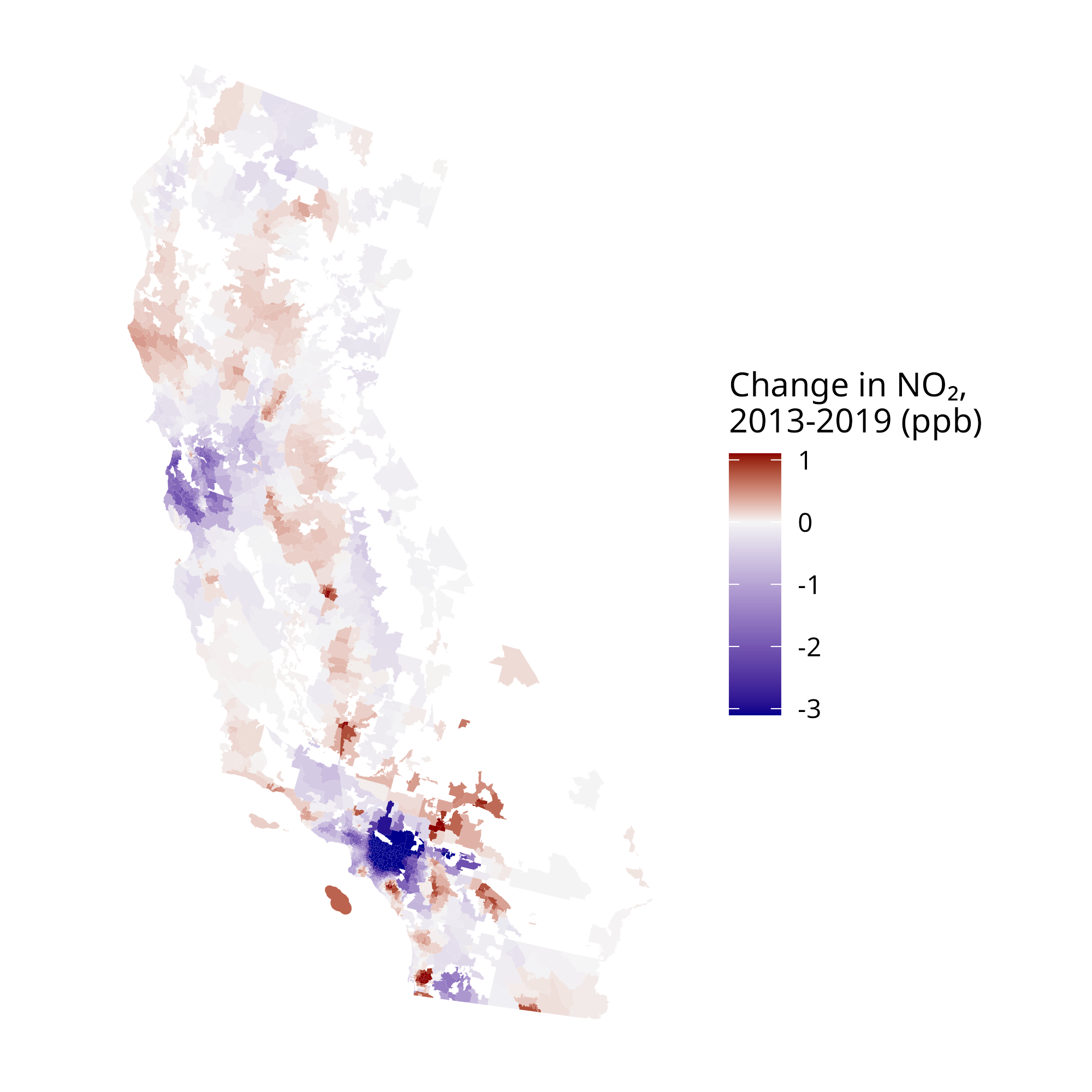}
  \caption{NO\textsubscript{2} air pollution in California.}
  \label{fig:NO2map}
\end{subfigure}%
\begin{subfigure}{.5\textwidth}
  \centering
  \includegraphics[scale=0.6]{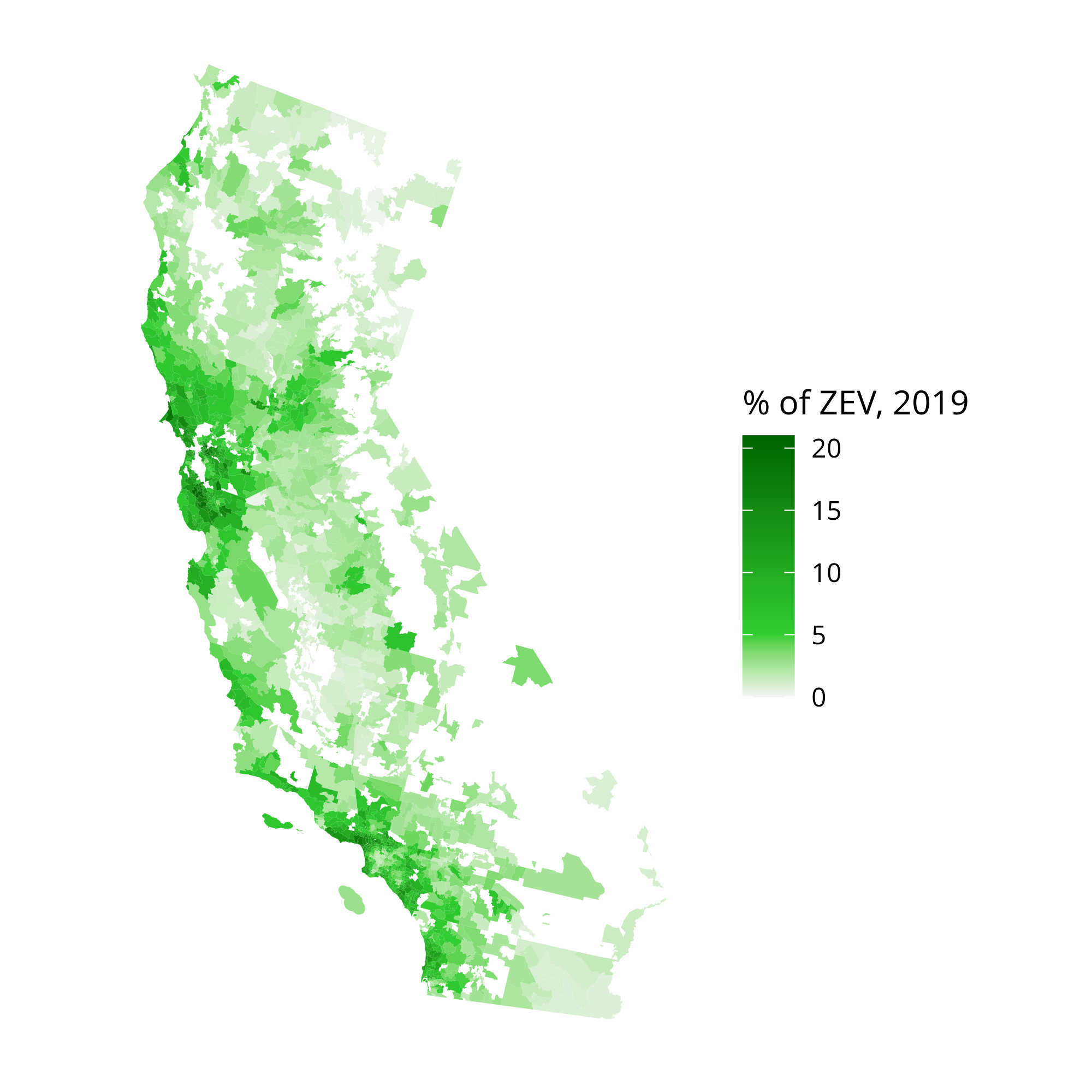}
  \caption{ZEV prevalence in California.}
  \label{fig:ZEVmap}
\end{subfigure}
\caption{Spatial distribution of outcome (NO\textsubscript{2}) and exposure
  (ZEV) in the data example.}
\label{fig:maps}
\end{figure}

Here, we include summary statistics for the dataset used to analyze the effect
of zero-emission vehicle uptake on NO\textsubscript{2} air pollution in
California. Figure~\ref{fig:maps}
depicts the spatial distribution of the exposure and outcome across the study
region of California. Table~\ref{table:summary-stats}
displays basic summary statistics for the exposure (Change in
NO\textsubscript{2}, 2013-2019), outcome (Percentage of ZEVs, 2019), and
confounders controlled for in the analysis described in the main manuscript.

\begin{table}[!htbp] \centering 
\caption{Semi-synthetic comparison, large exposure effect ($\mu_A \approx 8.4$)}
  \label{table:semisynthetic-big-mean}
\begin{tabular}{@{\extracolsep{5pt}} ccccccc} 
\\[-1.8ex]\hline 
\hline \\[-1.8ex] 
 & Method & Learner & Bias (\%) & Variance & Coverage (\%) & CI Width \\ 
\hline \\[-1.8ex] 
& \multirow{2}{*}{Network TMLE} & Correct GLM & $1.54$ & $1.56$ & $94.0$ & $4.88$ \\ 
& & Super Learner & $2.40$ & $1.58$ & $94.5$ & $4.91$ \\ 
& Classical TMLE & \multirow{2}{*}{Correct GLM} & $29.10$ & $1.00$ & $32.0$ & $3.92$ \\ 
& Linear Regression & & $29.53$ & $1.00$ & $31.0$ & $3.92$ \\ 
\hline \\[-1.8ex] 
\end{tabular} 
\end{table} 

\begin{table}[!htbp] \centering 
\caption{Semi-synthetic comparison, small exposure effect ($\mu_A \approx 0.4$)}
  \label{table:semisynthetic-small-mean}
\begin{tabular}{@{\extracolsep{5pt}} ccccccc} 
\\[-1.8ex]\hline 
\hline \\[-1.8ex] 
 & Method & Learner & Bias (\%) & Variance & Coverage (\%) & CI Width \\ 
\hline \\[-1.8ex] 
& \multirow{2}{*}{Network TMLE} & Correct GLM & $0.27$ & $1.56$ & $94.0$ & $4.88$ \\ 
&  & Super Learner & $0.32$ & $1.64$ & $94.5$ & $5.00$ \\ 
& Classical TMLE & \multirow{2}{*}{Correct GLM} & $5.29$ & $5.15$ & $91.8$ & $8.90$ \\ 
& Linear Regression & & $5.26$ & $5.16$ & $91.0$ & $8.91$ \\ 
\hline \\[-1.8ex] 
\end{tabular}
\end{table}

\begin{table}[H] \centering
  \caption{Summary statistics of ZEV-NO\textsubscript{2} data, including confounders ($n = 1652$)}
  \label{table:summary-stats}
\begin{tabular}{@{\extracolsep{5pt}}lcccc} 
\\[-1.8ex]\hline 
\hline \\[-1.8ex] 
Statistic & \multicolumn{1}{c}{Mean} & \multicolumn{1}{c}{Median} & \multicolumn{1}{c}{25\textsuperscript{th} pctl.} & \multicolumn{1}{c}{75\textsuperscript{th} pctl.} \\ 
\hline \\[-1.8ex] 
Change in NO\textsubscript{2}, 2013-2019 (ppb) & $-$0.6 & $-$0.1 & $-$0.9 & 0.1 \\ 
\% ZEV (of registered vehicles) , 2019 & 5.4 & 4.2 & 2.3 & 7.6 \\ 
\% ZEV (of registered vehicles), 2013 & 2.7 & 2.0 & 1.1 & 3.6 \\ 
Population & 23,753 & 19,505 & 2,485 & 38,478 \\ 
Median income (\$) & 76,944 & 69,156 & 51,299 & 95,552 \\ 
Median home value (\$) & 551,847 & 448,650 & 277,150 & 704,850 \\ 
Median age (years) & 40.9 & 39.2 & 34.4 & 46.1 \\ 
\% pop. college educated & 32.3 & 27.7 & 16.3 & 46.2 \\ 
\% pop. high school educated & 84.7 & 89.0 & 78.6 & 94.4 \\ 
\% pop. white & 69.4 & 73.7 & 55.7 & 85.8 \\ 
\% pop. in poverty & 10.1 & 7.5 & 3.8 & 14.1 \\ 
\% of homes owner-occupied & 59.8 & 62.6 & 47.6 & 74.1 \\ 
\% pop. who take automobile to work & 72.8 & 76.2 & 69.5 & 80.2 \\ 
\% pop. who take public transit to work & 3.5 & 1.2 & 0.0 & 3.5 \\ 
\% pop. who work from home & 8.0 & 6.0 & 3.8 & 9.7 \\ 
Industrial employment (jobs/acre) & 0.6 & 0.1 & 0.01 & 0.5 \\ 
Road density (per acre) & 11.1 & 6.8 & 2.4 & 19.8 \\ 
Public transit frequency (peak, per hour) & 5.4 & 1.0 & 0.1 & 4.9 \\ 
Walkability index & 8.7 & 7.6 & 5.1 & 12.3 \\ 
Network degree & 505.4 & 594 & 227.5 & 742 \\ 
\hline \\[-1.8ex]
\end{tabular}
\end{table}

\pagebreak
\bibliography{refs}

%% file: _preamble.tex
\usepackage{float}
\usepackage{graphicx}
\usepackage{comment}
\graphicspath{ {figures/} }
\usepackage{svg}
\usepackage[english]{babel}
\usepackage{dsfont}
\usepackage[OT1]{fontenc}
\usepackage[utf8]{inputenc}
\usepackage{lmodern}
\usepackage{tikz}
\usetikzlibrary{arrows}
\usepackage[normalem]{ulem}
\usepackage[symbol]{footmisc}

\usepackage{refcount,nameref,zref-xr,zref-user} % nameref before zref-xr
\zxrsetup{toltxlabel} % allows to use the \ref command as usual
\usepackage{url}
\usepackage{cancel}

\let\href\undefined
\usepackage{mathtools}
\usepackage{amsfonts}
\usepackage{amssymb}
\usepackage{bbm}
\usepackage{amsthm}
\usepackage{amsxtra}
\usepackage{amsmath}
\usepackage{commath}
\usepackage{times}
\usepackage{bm}
\usepackage{mathrsfs}

\newcommand{\abstracttext}{
    Modified treatment policies are a widely applicable class of interventions
    useful for studying the causal effects of continuous exposures. Approaches
    to evaluating their causal effects assume no interference, meaning that
    such effects cannot be learned from data in settings where the exposure of
    one unit affects the outcomes of others, as is common in spatial or network
    data. We introduce a new class of intervention---induced modified treatment
    policies---which we show identify such causal effects in the presence of
    network interference. Building on recent developments for causal inference
    in networks, we provide flexible, semi-parametric efficient estimators of
    the statistical estimand. Numerical experiments demonstrate that an induced
    modified treatment policy can eliminate the causal, or identification, bias
    that results from network interference. We use the methodology developed to
    evaluate the effect of zero-emission vehicle uptake on air pollution in
    California, strengthening prior evidence.
}

\makeatletter%
\@ifclassloaded{oup-authoring-template}{%
  \theoremstyle{thmstyleone}%
    \newtheorem{theorem}{Theorem}%  meant for continuous numbers
    \theoremstyle{thmstyletwo}%
    \newtheorem{example}{Example}%
    \newtheorem{remark}{Remark}%
    \theoremstyle{thmstylethree}%
    \newtheorem{definition}{Definition}
    \usepackage[doublespacing]{setspace}
    \usepackage[fontsize=12pt]{fontsize}

    \newtheorem{assumption}{Assumption}%
    \newtheorem{lemma}{Lemma}

    \bibliographystyle{abbrvnat}

  \newcommand{\authorlist}{}

}{%
  \usepackage{arxiv}
  \usepackage{standalone}
  \usepackage{multirow}
  \usepackage{amsthm}
  \usepackage[round]{natbib}
  \usepackage[inline,shortlabels]{enumitem}
  \bibliographystyle{biometrika}
  \newtheorem{theorem}{Theorem}

  {\theoremstyle{definition}\newtheorem{assumption}{}}
  {\theoremstyle{definition}\newtheorem{example}{Example}[section]}

  \newcommand{\authorlist}{
    Salvador V.~Balkus \\
    Department of Biostatistics, \\
    Harvard T.H.~Chan School of Public Health\\
    \texttt{sbalkus@g.harvard.edu}
    \And
    Scott W.~Delaney \\
    Department of Environmental Health, \\
    Harvard T.H.~Chan School of Public Health\\
    \texttt{sdelaney@mail.harvard.edu}
    \And
    Nima S.~Hejazi \\
    Department of Biostatistics, \\
    Harvard T.H.~Chan School of Public Health\\
    \texttt{nhejazi@hsph.harvard.edu}
  }
  \newcommand{\makeabstract}{
  \begin{abstract}
    \abstracttext
  \end{abstract}
  }
}
\makeatother

\newcommand{\titlepaper}{The causal effects of modified treatment
policies under network interference}

\AtEndDocument{\refstepcounter{theorem}\label{finalthm}}
\AtEndDocument{\refstepcounter{equation}\label{finaleq}}
\AtEndDocument{\refstepcounter{lemma}\label{finallemma}}

\newcommand{\E}{\mathbb{E}}

\newcommand{\M}{\mathcal{M}}

\newcommand{\logit}{\text{logit}}

\newcommand{\Pf}{\mathsf{P}}
\newcommand{\I}{\mathbbm{I}}

\newcommand{\indep}{\mbox{$\perp\!\!\!\perp$}}

\newcommand{\Ob}{\mathbf{O}}
\newcommand{\Lb}{\mathbf{L}}
\newcommand{\Ab}{\mathbf{A}}
\newcommand{\Yb}{\mathbf{Y}}
\newcommand{\Fb}{\mathbf{F}}
\newcommand{\A}{\mathcal{A}}
\newcommand{\Li}{\mathcal{L}}
\newcommand{\ab}{\mathbf{a}}

\newcommand{\lb}{\mathbf{l}}

\newcommand{\jrssb}[1]{\textcolor{black}{#1}}